\documentclass{aa}

\usepackage{txfonts}
\usepackage{graphicx}
\usepackage{amssymb}
\usepackage{CJKutf8}
\usepackage{soul}

\usepackage{natbib}
\usepackage{xcolor}
\usepackage{url}

\newcommand{\rom}[1]{\uppercase\expandafter{\romannumeral #1\relax}}
\DeclareMathOperator\erf{erf}

\begin{document}

   \title{Reconstructing AGN X-ray spectral parameter distributions with Bayesian methods. \newline 
\rom{2}. Population inference}

   \author{Lingsong Ge
          %\begin{CJK*}{UTF8}{gkai}
          %(葛泠松)
          %\end{CJK*}
          \inst{1}
          \and
          St\'ephane Paltani\inst{1}
          \and
          Dominique Eckert\inst{1}
          \and
          Mara Salvato\inst{2}
          }
          
   \institute{Department of Astronomy,  University of Geneva, ch. d'Écogia 16, CH-1290 Versoix, Switzerland\\ \email{lingsong.ge@unige.ch}
              \and
              Max-Planck-Institut für extraterrestrische Physik, Gießenbachstraße 1, D-85748 Garching, Germany
             }

   \date{Received xx, 2021; accepted xx, 202x}

% \abstract{}{}{}{}{} 
% 5 {} token are mandatory
 
  \abstract{We present a new Bayesian method for reconstructing the parent distributions of X-ray spectral parameters of active galactic nuclei (AGN) in large surveys. The method uses the probability distribution function (PDF) of posteriors obtained by fitting a consistent physical model to each object with a Bayesian method. The PDFs are often broadly distributed and may present systematic biases, such that naive point estimators or even some standard parametric modeling are not sufficient to reconstruct the parent population without obvious bias. Our method uses a transfer function computed from a large realistic simulation with the same selection as in the actual sample to redistribute the stacked PDF and then forward-fit a nonparametric model to it in a Bayesian way, so that the biases in the PDFs are properly taken into account. In this way, we are able to accurately reconstruct the parent distributions. We apply our spectral fitting and population inference methods to the \emph{XMM}-COSMOS survey as a pilot study. For the 819 AGN detected in the COSMOS field, 663 (81\%) of which have spectroscopic redshifts (spec-z) and the others high-quality photometric redshifts (photo-z), we find prominent bi-modality with widely separated peaks in the distribution of the absorbing hydrogen column density ($N_\mathrm{H}$) and an indication that absorbed AGN have harder photon indices. A clear decreasing trend of the absorbed AGN fraction versus the intrinsic 2--10\,keV luminosity is observed, but there is no clear evolution in the absorbed fraction with redshift. Our method is designed to be readily applicable to large AGN samples such as the XXL survey, and eventually eROSITA.}

   
   \titlerunning{Reconstructing AGN X-ray spectral parameter distributions (\rom{2})}
   \authorrunning{L. Ge et al.}

   \maketitle
%
%________________________________________________________________

\section{Introduction}

Active galactic nuclei (AGN) occupy the innermost regions of active galaxies, where supermassive black holes (SMBHs) accrete large amounts of material from their surroundings, converting a large fraction of the accretion energy \citep[$\approx$5--42\% depending on the spin of the SMBH;][]{Kerr1963,Shapiro1983} into radiation throughout most of the electromagnetic spectrum. Being some of the most luminous steady sources in the Universe, AGN are detectable out to high redshifts \citep{Vito2018}. All galaxies are thought to host a SMBH in their center \citep{Magorrian1998}, and the majority of galaxies are believed to have gone through an AGN phase in the past \citep{Marconi2004,Merloni2008}. For this reason, determining the history of accretion onto SMBHs and their properties is key to understanding galaxy evolution. The evolution of AGN activity also allows us to probe the origin of the tight relationship between SMBH mass and host galaxy properties \citep{Magorrian1998,Ferrarese2000,Marconi2003,Haring2004,Gultekin2009,Kormendy2013}, and their similar growth history \citep{Hopkins2006,Silverman2008,Aird2015}.

To see the full picture of the evolution of SMBHs and its impact on the host galaxies, a complete demographic study of AGN is required. However, this is a challenging task as most AGN \citep[$\approx 90$\%;][]{Hickox2018} are obscured in the optical and ultraviolet (UV) by the intervening gas and dust, which is thought to be mostly contained in a dusty torus \citep{Krolik1988}. A typical AGN consists in particular of the accretion disk, the broad-line region (BLR), and the dusty torus. The BLR is a dust-free region close to the central SMBH on subparsec scales and produces broad emission lines (full width at half maximum $\ge 2000$\,km s$^{-1}$) in the optical/UV band; they are used to classify AGN as type-1 or type-2 depending on their presence or absence in the optical spectrum \citep{Antonucci1993}. The dusty torus is a geometrically thick toroidal structure well outside of the accretion disk and the BLR on a scale of several parsecs, which largely obscures the direct emission from the central region of the AGN when we observe through it \citep{Krolik1988,Urry1995}. Therefore, optical surveys are heavily biased toward bright type-1 AGN and very inefficient at detecting obscured AGN, even though the AGN spectral energy distribution usually peaks in the optical/UV through thermal emission from the accretion disk \citep{Shakura1973}. Mid-infrared (mid-IR) AGN surveys are less affected by the obscuring torus compared to the optical because they are sensitive to the reprocessed emission from the torus that is observed in this band \citep{Sanders1989}. However, in cases where the host galaxies are also dusty, the AGN signal is diluted, restricting the mid-IR-selected sample to luminous sources \citep{Donley2008}. 

Among the many different AGN detection methods, X-ray observations are one of the most complete and least biased ways to select AGN \citep{Brandt2015} thanks to the ubiquitous X-ray emission in AGN \citep{Avni1986,Brandt2000,Gibson2008}, which always outshines that of the host galaxy \citep{Maccacaro1988}, and to the strong penetrating power of X-rays through gas and dust \citep{Wilms2000}. Both wide and deep X-ray surveys have been conducted to study the evolution and population properties of AGN up to redshifts greater than 3 \citep[e.g.,][and many others]{Giacconi2002,Hasinger2007,elvis2009,Pierre2016,Luo2017}, taking advantage of the efficiency and completeness of X-ray selection, especially in the hard X-ray band (i.e., above 2\,keV).

The intrinsic X-ray emission of AGN is thought to be produced by the inverse Comptonization of disk photons in a hot and optically thin corona in the vicinity of the SMBH \citep{Blandford1990,Zdziarski1995,Zdziarski1996,Krolik1999}. This emission appears in the X-ray spectrum as a power law, namely \(N(E) \propto E^{-\Gamma}\), with $E$ the energy of the photon and $\Gamma$ the photon index. This parameter is an important indicator of the AGN growth rate: previous studies have shown that the 2--10\,keV intrinsic photon index is correlated with the Eddington ratio \citep{Pounds1995,Risaliti2009,Jin2012,Brightman2013}, the ratio of the AGN bolometric luminosity to the Eddington luminosity ($L_\mathrm{Edd}$), the maximum theoretical luminosity of an accreting object. This correlation may also be used to estimate the SMBH mass through the dependence of $L_\mathrm{Edd}$ \citep{Shemmer2008}.

A reliable measurement of the intrinsic photon index requires a careful correction for the X-ray absorption due to the hydrogen column density, $N_\mathrm{H}$, along the line of sight since these two parameters are degenerate \citep[see Fig.~12 in][submitted; hereafter Paper~\rom{1}]{Ge2021}. Therefore, properly determining $N_\mathrm{H}$ is essential for constraining the distribution of $\Gamma$ and inferring the accretion history of SMBHs, as well as other demographic properties, such as the evolution of the intrinsic X-ray luminosities \citep[$L_\mathrm{X}$;][]{Ueda2014,Aird2015,Buchner2015}. In addition, the $N_\mathrm{H}$ distribution is an important indicator of the structure of AGN and is a necessary ingredient for understanding the physical properties of the nuclear regions. For example, the unified AGN model \citep{Antonucci1993,Urry1995} successfully explains the different types of AGN classified at optical wavelengths based on the orientation of the torus with respect to the line of sight. However, more recent surveys found clear evidence that the absorbed AGN fraction, defined as the fraction of sources with $N_\mathrm{H}$ larger than $10^{22}$\,cm$^{-2}$, decreases with increasing intrinsic X-ray luminosity \citep{Ueda2003,Hasinger2008,Brusa2010,Merloni2014,Buchner2015}, which points toward the existence of intrinsic differences between absorbed and unabsorbed AGN. To explore the physics behind this dependence, \citet{Ricci2017b} analyzed a hard X-ray-selected (14--195\,keV) sample from the survey performed by the Burst Alert Telescope \citep[BAT;][]{Barthelmy2005} on the \emph{Swift} X-ray observatory \citep{Gehrels2004}, showing that the absorbed AGN fraction depends on $L_\mathrm{Edd}$, which suggests that the obscuration of AGN is mostly driven by the mass-normalized accretion rate.

Besides the primary power law, X-ray spectra of AGN also exhibit a number of secondary components, including an excess of soft X-ray emission below 1--2\,keV \citep{Arnaud1985,Singh1985}, a reflected component from the inner region of the AGN by the circumnuclear material \citep{Pounds1990,Nandra1994}, a scattering component by warm, fully ionized gas relatively far away from the center \citep{Bianchi2006,Guainazzi2007}, and thermal emission from diffuse plasma in the host galaxy \citep{Iwasawa2011}. This makes the total spectrum very complex, and such complexity in turn makes the extraction of the main spectral parameters (e.g., $N_\mathrm{H}$, $\Gamma$, and $L_\mathrm{X}$) without strong bias a non-trivial task, especially when considering large samples of sources with low signal-to-noise ratios (S/N), which is typical of large surveys. This point is addressed in Paper~\rom{1}, which presents our method for AGN X-ray spectral fitting and its validation using simulations. 

In this paper we developed a new Bayesian method for reconstructing the distributions of spectral parameters of the parent AGN population based on the probability distribution functions (PDFs) that are obtained from the spectral fitting step. We applied this method to the \emph{XMM}-COSMOS survey of the COSMOS field \citep{Hasinger2007} performed with \emph{XMM-Newton} \citep{Jansen2001}. The details of data reduction and source detection in the \emph{XMM}-COSMOS survey are introduced before presenting the method (Sect.~\ref{cosmos}). We created a simulated sample to support our analysis in Sect.~\ref{simulation}, and in Sect.~\ref{method} we reconstructed the parent distributions of the main spectral parameters. Since the PDFs from fits to low-S/N data can be biased due to the inhomogeneous sensitivity of the likelihood for some parameters (e.g., $N_\mathrm{H}$), we introduced a transfer function to calibrate this bias in the PDF and perform a reliable population inference (see Sect.~\ref{new method}). We validated our method using extensive simulations (Sect.~\ref{Validation of FFPI}) and applied it to the \emph{XMM}-COSMOS survey as a pilot study to reconstruct the distributions of $N_\mathrm{H}$ and $\Gamma$, as well as the absorbed fraction versus $L_\mathrm{X}$ and redshift (Sect.~\ref{results}). Throughout the paper we assumed the WMAP9 $\Lambda$ cold dark matter cosmology: $H_0=69.3$ km/s/Mpc, $\Omega_\mathrm{\Lambda}=0.72$, $\Omega_\mathrm{m}=0.28$,  and $\sigma_\mathrm{8}=0.82$ \citep{Hinshaw2013}.

%_________________________________________________________________
\section{The \emph{XMM}-COSMOS survey}
\label{cosmos}

The \emph{XMM}-COSMOS survey centered at RA +150.11916667 (10:00:28.600), Dec = +2.20583333 (+02:12:21.00) \citep{Hasinger2007,Scoville2007} consists of $\sim$60 ks observations of the 2 deg$^2$ COSMOS field with \textit{XMM-Newton}. 1848 point-like sources have been detected in the soft and/or hard X-rays \citep{Cappelluti2009}. This field is an excellent test bed for our new method, because it has been studied extensively in various multiwavelength surveys \citep{Brusa2010}, thanks to which almost all sources have either spectroscopic redshifts (spec-z) or high-quality photometric redshifts \citep[photo-z;][]{Salvato2009,Salvato2011,Laigle2016,Weaver2021}.

\subsection{Data preparation}
We retrieved observation data files (ODFs) in the COSMOS field for both the EPIC pn \citep{Struder2001} and EPIC MOS \citep{Turner2001} cameras from the \textit{XMM-Newton} Science Archive and found 56 pointings (see the detailed observation log in Appendix~\ref{observation list}). Starting from the ODFs, we reduced the observation data using the data analysis pipeline described in detail in Sect.~2.2 of \citet{Ghirardini2019}, which is based on the \textit{XMM-Newton} Science Analysis Software v13.5 (XMMSAS). \texttt{emchain} and \texttt{epchain} were used to extract calibrated event files. The \texttt{mos-filter} and \texttt{pn-filter} executables were run to filter out time periods affected by soft-proton flares. We also measured count rates in unexposed corners of both instruments to compare them with those inside the field of view (FOV) of the telescope, in order to estimate the contamination of residual soft protons in the spectrum.

Count images were extracted from the pn, MOS1, and MOS2 detectors in nine bins: [0.6,\,0.9], [0.9,\,1.3], [1.3,\,1.9], [1.9,\,2.8], [2.8,\,4.2], [4.2,\,6.3], [6.3,\,7.2], [7.2,\,9.2],  and [9.2,\,12]\,keV. The 0.4--0.6\,keV band was ignored because of the unstable energy gain of the pn and MOS cameras at low energy, which renders the calibration uncertain. The sizes of the bins were selected to be close to a uniform distribution in log space, except for two of them, which were set to 1.3--1.9\,keV and 7.2--9.2\,keV because of the Al\,K${\alpha}$ and Si\,K${\alpha}$ fluorescence emission lines at 1.5 and 1.7\,keV for both pn and MOS, and those of Cu\,K${\alpha}$, Ni\,K${\alpha}$ and Zn\,K${\alpha}$ around 8\,keV for pn (see \textit{XMM-Newton} Users Handbook \footnote{\url{https://xmm-tools.cosmos.esa.int/external/xmm_user_support/documentation/uhb/epicintbkgd.html}}). We used \texttt{eexpmap} to compute exposure maps for the three detectors independently, taking into account vignetting. For source detection, we extracted images and exposure maps in the 2--7\,keV band, which allowed us to be sensitive to absorbed sources that are undetected at low energy. At the same time, restricting the upper bound of the band to 7\,keV enhanced the sensitivity of the map, as the ratio of effective area to background is unfavorable beyond 7\,keV. The total image was created by summing up the count maps of all detectors, and the corresponding exposure map was computed by summing up the MOS exposure maps and those of pn multiplied by the ratio of pn to MOS effective area averaged over the energy band, which for a power law source with a photon index of 2.0 is 2.76. In the end we combined images and exposure maps of all pointings to obtain the mosaics of the total field, which allowed us to make full use of the regions located on overlapping pointings.

\subsection{Source detection}
\label{source detection}

For point source detection we used the XMMSAS tool \texttt{ewavelet} on the 2--7\,keV mosaic. The wavelet scale was set to 2--4 pixels and the significance threshold to 5. A total of 860 sources were detected. The much lower number of sources compared to the analysis of \citet{Brusa2010} is due to the selection in the hard band only, while \citet{Brusa2010} also used the 0.5--2\,keV band, which is intrinsically more sensitive but highly biased toward unabsorbed objects. We cross-matched our source list with the \emph{Chandra} COSMOS Legacy survey catalog that includes 4016 X-ray sources, 97\% of which are identified in the optical and in the near-infrared \citep{Civano2016,Marchesi2016}. We found a match with the \emph{Chandra} catalog with a separation smaller than 10 arcsec for 828 sources. Five of the remaining unmatched 32 sources were within 20 arcsec of objects in the cluster catalog of \citet{Gozaliasl2019}. There were also 13 sources outside of the field of the \emph{Chandra} COSMOS Legacy survey. Therefore, we ended up with 14 unmatched sources (1.6\%), half of which are close to the boundary of the \emph{XMM}-COSMOS survey, so that their positions could be inaccurate. Among the 828 matched sources, there were 48 sources with two counterparts in the \emph{Chandra} catalog. To select the real counterpart and get the correct redshift information, we first looked at their fluxes and relative distances to the \textit{XMM-Newton} source. If one of the \emph{Chandra} counterparts was three times brighter or 5 arcsec closer than the other one, we selected it as the real counterpart. In this way we identified the counterparts of 21 sources. For the remaining 27 we kept those having only one \emph{Chandra} counterpart within 3 arcsec, and ignored sources that had both counterparts either within 3 arcsec or farther away than 3 arcsec, since it is difficult to distinguish between them. Our final sample contains 819 point sources matched with a single \emph{Chandra} counterpart in the 2--7\,keV band, 663 (81\%) of which have measured spectroscopic redshifts. We assumed that the missing objects are purely random, so that distributions are unaffected by the exclusion of these sources.

The spectra of these sources were extracted directly from the 10-band mosaics using our C++ aperture photometry tool \texttt{xphot}\,\footnote{\url{https://github.com/domeckert/xphot}}, as described in Paper~\rom{1}. The radius of the source region was fixed to 30 arcsec, and the background region was an annulus from 60 arcsec to 90 arcsec away from the source center, to avoid remaining contamination from the source. The areas covered by neighboring sources (with 30-arcsecond-radius aperture) in the catalog were masked when extracting the background spectrum. In this paper we neglected the impact of photons smeared more than 30 arcsec away from the source position by the \emph{XMM-Newton} point spread function (PSF), as except for a $\sim10$\% underestimation of the luminosities the effect was found to be small (see Appendix \ref{sec:psf}). A correction for photon loss by the PSF will be included when applying our technique to future surveys.

\section{Simulations}
\label{simulation}

An accurate knowledge of the survey selection function is necessary to reconstruct the true properties of the parent population. To determine the selection function, we simulated a large set of sources with varying 2--7\,keV count rates. In addition, as will be explained in detail in Sect.~\ref{method}, these simulations will be used to determine the transfer function of our forward-fitting approach, allowing us to reconstruct the parameter distribution in the parent population.

\begin{figure}[tb]
\centering
\includegraphics[width=\columnwidth]{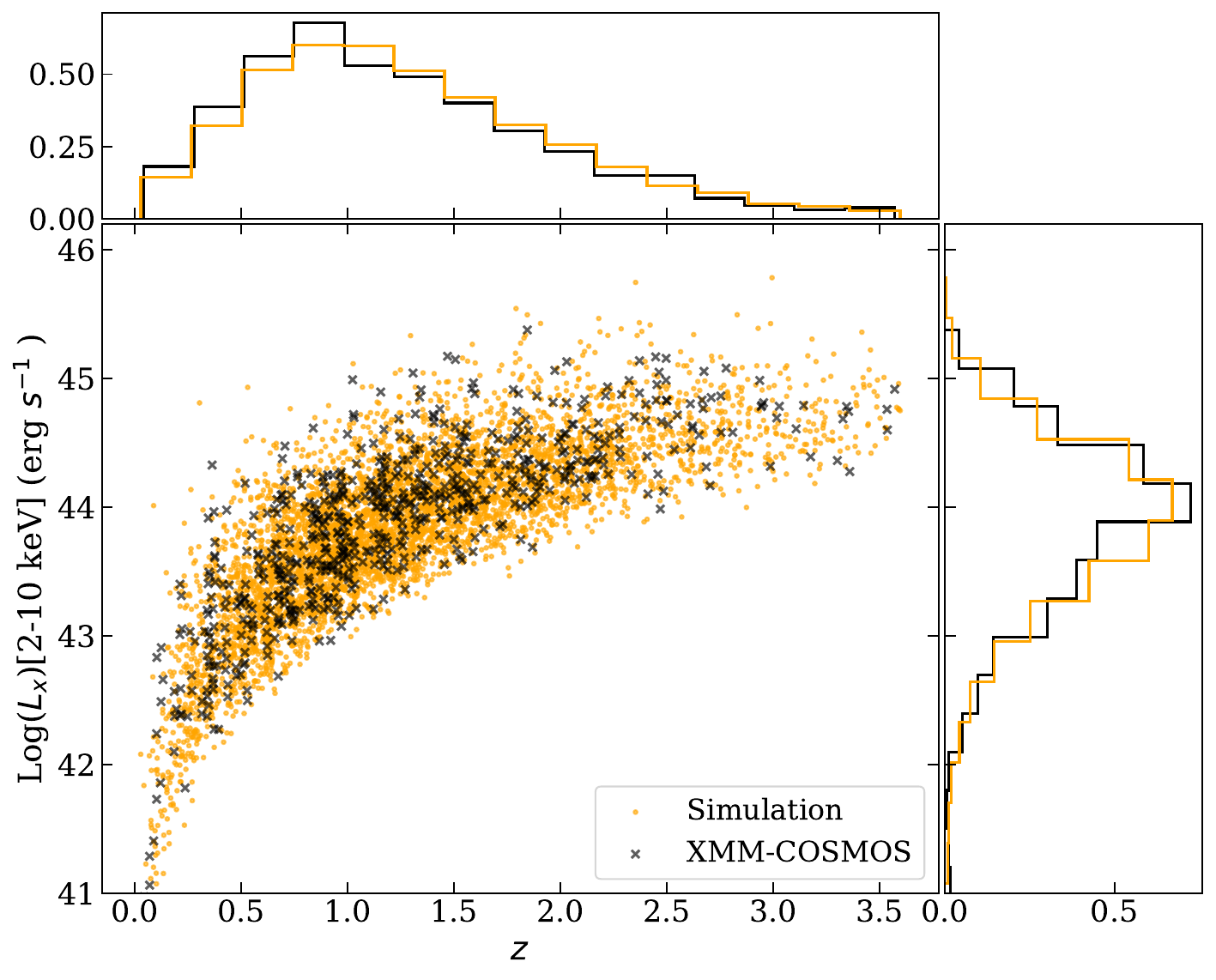}
\caption{Scatter plot of redshift versus intrinsic $L_{\mathrm{X}}$ in the 2--10\,keV band. The  histograms of these two parameters, normalized to have unit area, are plotted in the top and right subpanels. The simulated sample is plotted in orange, and the \emph{XMM}-COSMOS data sample is plotted in black. $L_\mathrm{X}$ is calculated using the median of the PDFs from the spectral analysis.}
\label{fig:simul_lx_z}
\end{figure}

\subsection{Simulation approach}
To characterize properly the selection function, we need to reproduce as precisely as possible the process that leads to the source detection. We first simulated the source count rate in the 2--7\,keV band using the full model described in Sect.~2.1 of Paper~\rom{1}, which consists of the primary power law, soft excess, reflection, scattering and emission from thermal plasma. Source spectral parameters are drawn from their priors, including a non-informative prior (flat in log-scale) for $N_\mathrm{H}$, and informative priors obtained from deep observations \citep{Ricci2017, Boissay2016} for parameters of secondary components (see Sect.~3.4 and Fig.~3 of Paper~\rom{1} for details). We drew however $\Gamma$ from a Gaussian distribution with $\mu = 1.95$ and $\sigma = 0.15$ \citep{Buchner2014}, instead of from an uninformative prior, to avoid generating completely unrealistic sources. We selected sources with redshifts and intrinsic X-ray luminosities according to the X-ray luminosity function (XLF) of \citet{Aird2015} given by the flexible double power law model \citep[FDPL in][]{Aird2015} with $0<z\le 3.6$ and $10^{41}\le L_\mathrm{X} \le 10^{46}$ erg\,s$^{-1}$, where 3.6 is the highest redshift we have in our \textit{XMM}-COSMOS data sample. 

Next, we directly added the simulated source counts at a randomly selected position in the real 2--7\,keV band detection mosaic and applied the same source detection algorithm again. The simulated-source position was determined after excluding a circular region of 10 arcsec radius around real sources detected in Sect.~\ref{source detection}. Although, ideally, real source regions are supposed to be included in the simulation, it is quite complex to distinguish the simulated source from the real one when only one source is detected; we therefore excluded sky areas that would result in such confusion (i.e., about twice the telescope's FOV-averaged half-energy width). Since the source regions defined in this way only occupy less than 1\% of the total area, the impact on our selection function is negligible. Source counts of simulated sources in the detection band and in all energy bands were computed using the local exposure time at the position in the exposure map, randomized by Poisson statistics and further spatially distributed using the PSF calculated using the XMMSAS tool \texttt{psfgen}, at 3\,keV and 8\,arcmin off-axis, which is close to the FOV-averaged and energy-weighted PSF in our survey. To avoid noticeably increasing the source density in the image, we only simulated 100 sources at a time. The detection band image was then processed with \texttt{ewavelet} in the same conditions as for the detection of real sources, and we compared the detected source list with the simulated list to assess the probability of detection. We used 10 arcsec as the matching distance, and the typical positional offset was within 3 arcsec. In each simulated image, only a fraction of around $5-10\%$ of the input sources were detected, since most of the simulated sources are intrinsically too faint because of the luminosity function. We point out that, at this stage, we are only interested in the simulated sources, so we discarded all other sources, real or spurious, that were present in the original data. We repeated the process until enough simulated sources were detected.

The spectra of the detected simulated sources were extracted using the same approach as the one we used for the \emph{XMM}-COSMOS data by applying the \texttt{xphot} executable to the two sets of nine mosaics, one for pn, and the other for the combined MOS instruments. We used the detected positions of the simulated sources to match as closely as possible the processing of the real sources. In this case, however, the source spectra were simply calculated by adding the Poisson distributed source counts to the spectral bins extracted from the source regions, which originally only contain background counts.

\subsection{Properties of the simulated sample}
\label{Simulated sample}

\begin{figure}[tb]
\centering
\includegraphics[width=\columnwidth]{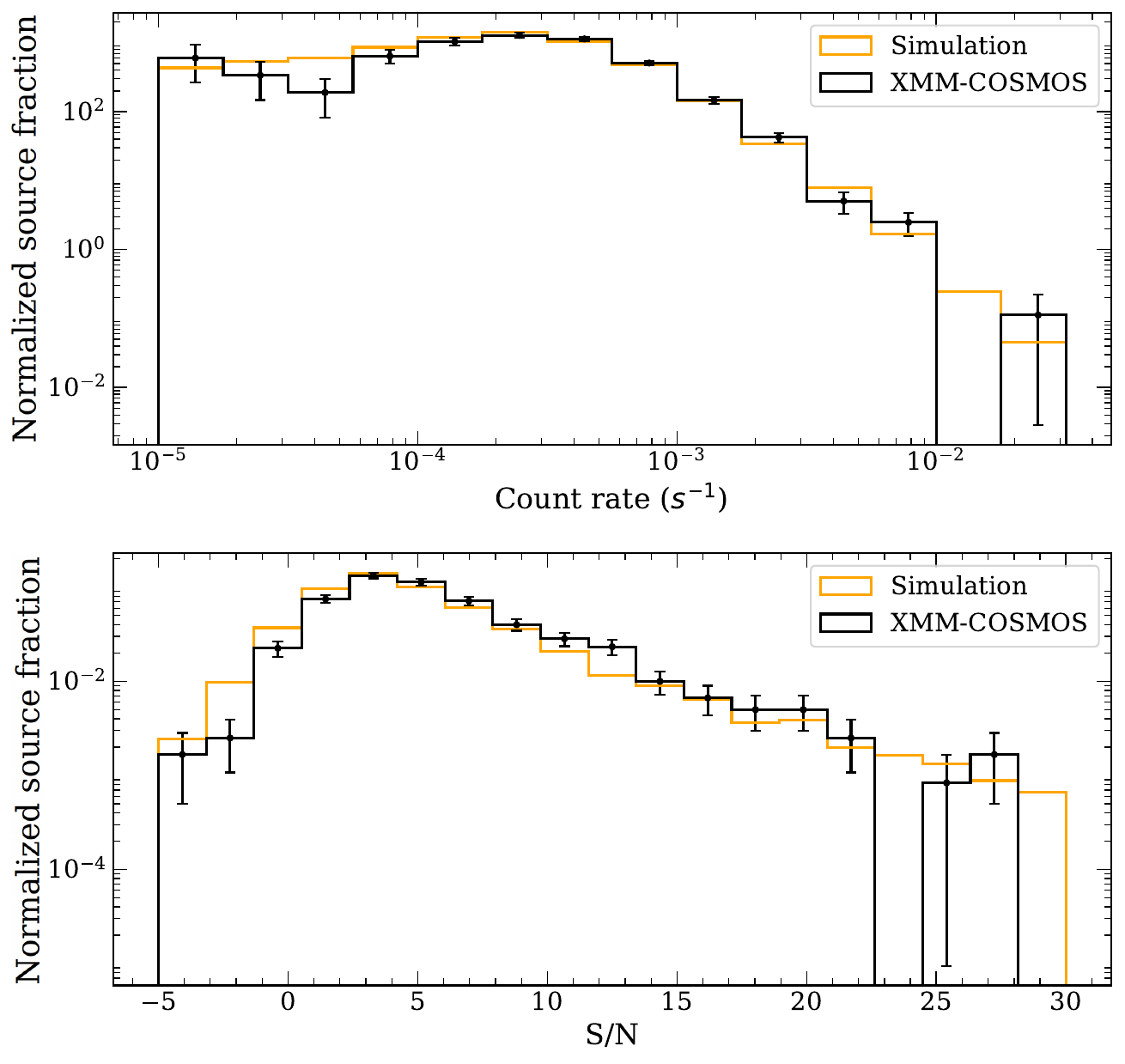}
\caption{Distributions of source count rates (top) and S/N (bottom) of the simulated sample (orange) and our \emph{XMM}-COSMOS sample (black) determined in the 2--7\,keV source detection band and normalized to unit area. 1$\sigma$ errors on the histograms of the \emph{XMM}-COSMOS sample are shown in black. The errors on the simulated sample are not presented since they are negligible in these plots.}
\label{fig:sample_cr_snr}
\end{figure}

We repeated the simulation process until we got over 10\,100 detected sources. Of them, 88 fall into boundaries with exposure from only one detector, so we excluded them as there isn't any such sources in our \emph{XMM}-COSMOS sample; 10\,030 sources finally remain in the simulated sample. The $L_\mathrm{X}$--$z$ distributions of the simulated sample and of the actual sample are shown in Fig.~\ref{fig:simul_lx_z}. The $L_\mathrm{X}$ of the real sources were determined using the medians of their PDFs (see Sect.~\ref{results}).

The distributions of (background subtracted) count rates and S/N in the source detection band of the two samples are presented in Fig.~\ref{fig:sample_cr_snr}. Sources with negative or too low count rates ($< 10^{-5}$\,s$^{-1}$) were excluded from the top panel of Fig.~\ref{fig:sample_cr_snr}, since the horizontal axis is in $\log$. S/N is defined as 
\begin{equation}
\label{eq:s/n}
\mathrm{S/N} = (N_\mathrm{s} - N_\mathrm{b}/r)/ \sqrt{N_\mathrm{s} + N_\mathrm{b}/r^2},  
\end{equation}
where $N_{\mathrm{s}}$ is the total count in the source region, $N_{\mathrm{b}}$ is that in the background region and $r$ is the ratio between background and source areas. We note that $r$ is not a fixed value, because it depends on the number and positions of other sources in the background region. We point out that \texttt{ewavelet} determines the S/N separately in a different manner, which explains why some sources are detected with formally negative count rates. The count rate distributions match very well, except for one bin at low count rate around $5 \times 10^{-5}$ s$^{-1}$. This is probably due to statistical fluctuations because of the much smaller sample size of the \emph{XMM}-COSMOS data sample, as shown by the error bars assuming they follow a binomial distribution. The two S/N distributions also match well, as they both peak around 4.5 and decrease in the same way below and above the peak. We note that the S/N for both samples extend below zero, because the background counts are estimated from the pixels in the annulus outside of the source region.

\subsection{Selection function}
\label{selection function}

\begin{figure}[tb]
\centering
\includegraphics[width=\columnwidth]{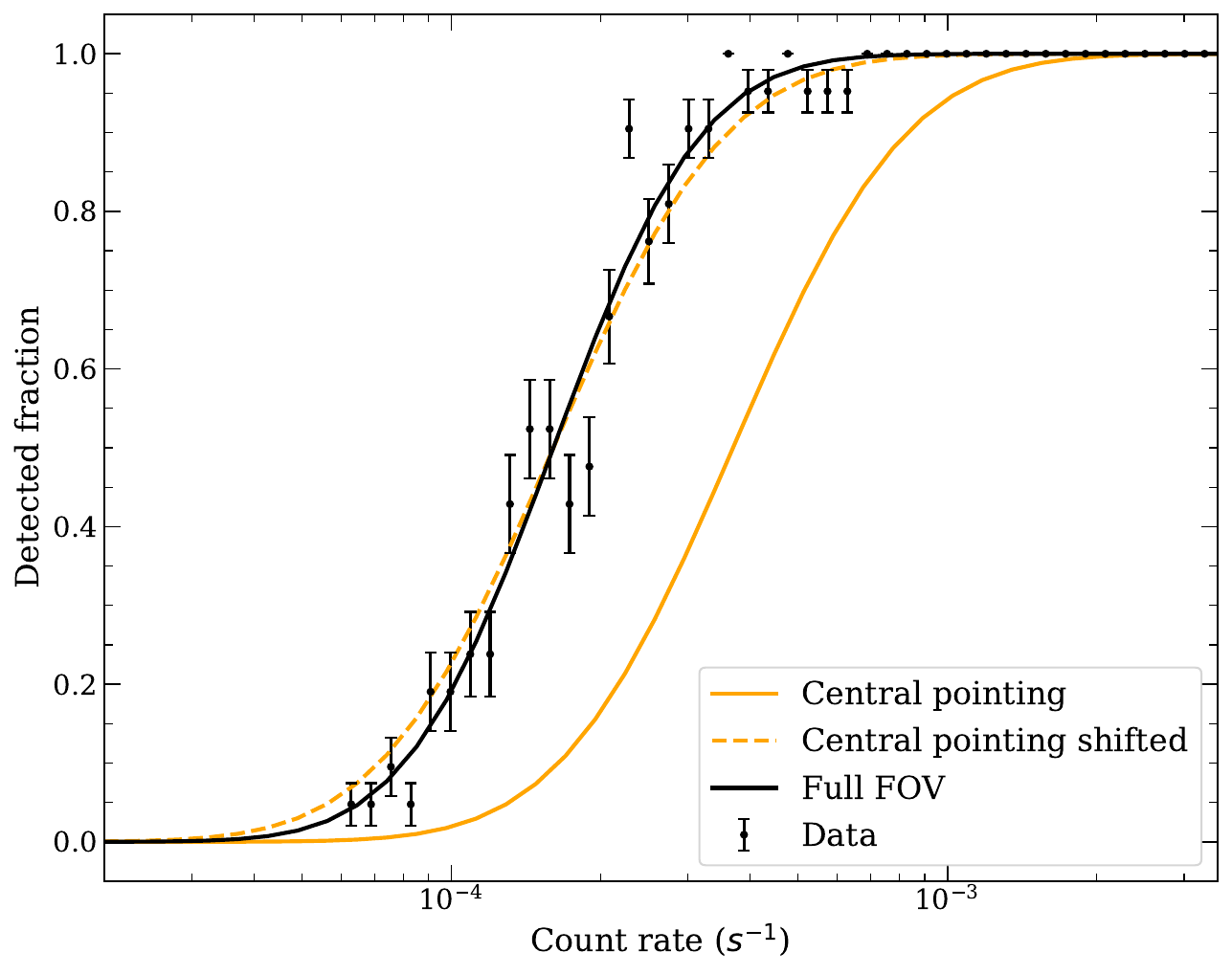}
\caption{Selection function of the full mosaic (solid black line) fitted to the detected fraction of our simulated sources (black). The selection function in the central pointing is shown as a solid orange line. The dashed orange line shows the central-pointing selection function shifted to the same depth as that of the full field (same parameter $b$ in Eq.~(\ref{eq:selfuc}) for comparison).}
\label{fig:selfuc}
\end{figure}

The selection function describes the probability of observing a source given its spectral properties. Since the detection is only a function of the number of counts in the detection band compared to the number of background counts, we parameterized the selection function as a function of the count rate $c$. The effect of varying background counts and exposure times was taken into account by the randomization of the position of the simulated sources in the simulated images. The parameterization was made using an error function as follows:
\begin{equation}
\label{eq:selfuc}
P(c|a,b) = 0.5 \erf\left( a (\log c +b)\right) + 0.5,
\end{equation}
where $a$ and $b$ are free parameters. We used a simple least-squares regression to fit this function to the detected fraction of simulated sources at different count rates, $c$. From the fit we obtained $a=3.0\pm0.3$, $b=3.79\pm0.02$, and the results are shown in Fig.~\ref{fig:selfuc}. We also computed the selection function in the same way on a single pointing (obsID:0203360801) in the central region of the field for comparison, taking into account the position-dependent PSF, whereas on the mosaic we used an FOV-averaged PSF. The selection fractions over the whole field and the fitted selection functions over the whole field and in the central pointing are shown in Fig.~\ref{fig:selfuc}. The best-fit $a$ and $b$ parameters for the central pointing are $2.6\pm0.3$ and $3.43\pm0.02$. To compare the shape of the two fitted functions, we shifted that of the single pointing, so that its $b$ parameter becomes the same as that of the full mosaic); we find that the shapes match quite well. Obviously, by combining the data of all the observations we reached much deeper depths compared to the single pointing, but we see that the effect of using a FOV-averaged PSF does not significantly affect the shape of the selection function.

The selection function obtained above was used as a prior in the spectral fitting process to penalize solutions that lead to an undetectable source in the 2--7\,keV detection band (see Sect.\,3.3 of Paper~\rom{1} for details).

%__________________________________________________________________
\section{Reconstruction of the parent population's properties}
\label{method}

To infer the distributions of the spectral parameters in the parent population using the PDFs of each source, the simple approach is to use a point estimator, for instance the median, or to stack the PDFs. The former can be a satisfactory approach only in cases where the PDFs are well constrained in a single peak, which requires good data quality and non-degenerate models. If the PDFs are widely distributed in the parameter space or are multimodal, using point estimators obviously cannot extract enough information from the full PDFs to perform a correct population inference. Stacking PDFs seems straightforward, but it is actually statistically meaningless, because probabilities of independent events are not additive, but multiplicative \citep{Malz2021}. So stacking PDFs can only give us a rough estimate of the parent population, and not a proper inference of the parent population. In addition, stacked PDFs contain uncertainties due to the measurement process, and provide too broad parent distributions \citep{Loredo2004}. We explored here two statistically correct methods to reconstruct the parent distributions, one of them being an entirely new approach, as far as we know.

\subsection{Parametric fitting method}
\label{parametric method}

The first method is to parameterize the parent distribution and then fit the model to the PDFs. We follow here J.~Speagle's approach\footnote{\url{https://github.com/joshspeagle/frankenz/blob/master/demos}: 5 -- Population Inference with Redshifts.ipynb}. Suppose our sample is $\{s\}$, and for each source $s_i$ we have an associated parameter estimate $\theta_i$ with PDF: $P(\theta_i|s_i)$. Since our goal is to reconstruct the parent distribution of parameter $\theta$ given $\{s\}$, which can be denoted as $\phi\equiv P(\theta|\{s\})$, we need a likelihood function for the PDFs, which we can write as $P(\{p_{\theta}\}|\phi)$. Now using Bayes theorem, we can write the posterior as
\begin{equation}
P(\phi|\{p_{\theta}\}) \propto P(\{p_{\theta}\}|\phi)P(\phi)
.\end{equation} Assuming the PDFs are independent, the likelihood can be rewritten as the product of all the PDFs marginalized over the parameter space as the following:

\begin{equation}
\label{eq:likelihood_1}
P(\{p_{\theta}\}|\phi) = \prod_{i} P(p_{\theta_i}|\phi) = \prod_{i} \int P(\theta_i|s_i)P(\theta|\{s\})d\theta
.\end{equation} Namely, the posterior probability of the population distribution is determined by maximizing its overlap with each PDF $P(\theta_i|\theta)$.

\citet{Loredo2004} includes another term in the likelihood function (Eq.~6), which asserts that no other sources are detected elsewhere in the parameter space, to account for the source uncertainties. We chose to simplify it and ignore this term for the parametric fitting method, because almost all the parameter space is covered by the PDFs of detected sources (especially for $\Gamma$, where this method is applied as presented below), and thus the additional term does not make much difference here.

We infer the parent distribution of the photon index, $\Gamma$, as an example of the application of this method. Since it has been observed that $\Gamma$ approximately follows a Gaussian distribution \citep{Nandra1994,Ueda2014,Ricci2017}, we parameterized its parent distribution with a Gaussian with only two free parameters: the mean, $\mu$, and the standard deviation, $\sigma$. We applied non-informative priors to the two parameters, including a uniform prior between 1.5 and 2.5 for $\mu$ and a Jeffreys prior between 0.01 and 1 for $\sigma$. With the likelihood function and prior we sampled the posterior using \texttt{MultiNest} \citep{Feroz2008,Feroz2009,Feroz2019}, which is an implementation of the nested-sampling algorithm \citep{Skilling2004}. \texttt{MultiNest} has the capability to sample multimodal posteriors, which is essential to analyze low-S/N data with complex models, since more common sampling approaches such as Markov chain Monte Carlo often have difficulties in exploring a multimodal parameter space. We tested this approach by constructing a subsample of 660 sources from the total simulated sample introduced in Sect.~\ref{Simulated sample} with a true $\Gamma$ distribution following a Gaussian with $\mu = 1.75$ and $\sigma = 0.15$. We selected this relatively small mean value of $\Gamma$ so that it is significantly different from the median of the prior of $\Gamma$, which is applied for the spectral fitting.

\begin{figure}[tb]
\centering
\includegraphics[width=\columnwidth]{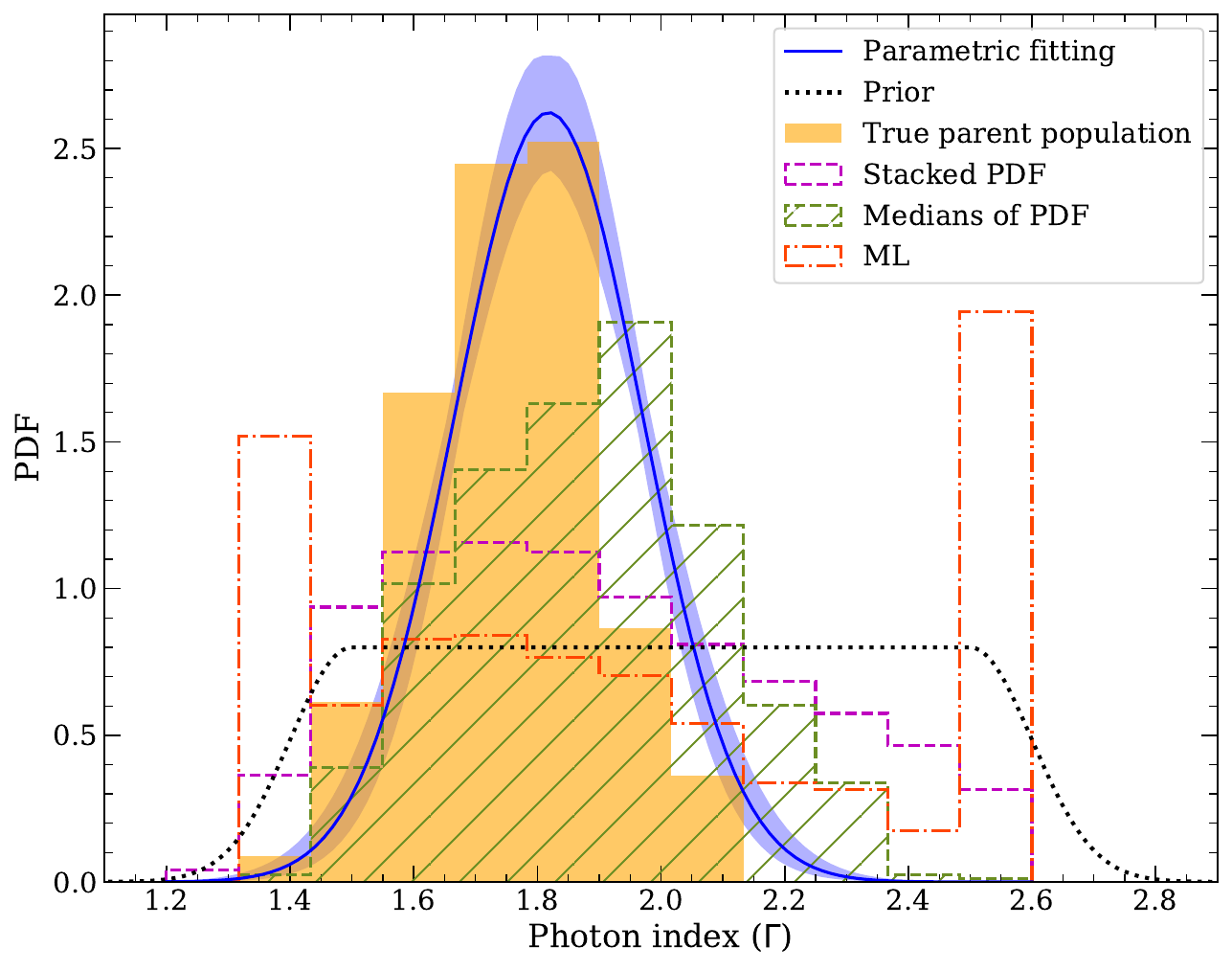}
\caption{Reconstructed parent distribution of $\Gamma$ of a simulated sample using the parametric fitting approach in blue (see Sect.~\ref{parametric method}), with the filled area showing the 1$\sigma$ credible interval. The true distribution is shown as the orange histogram, and the results using medians and stacked PDF are shown with dashed green (hatched) and magenta lines. The distribution of ML values is shown with a dash-dotted red line, and the dotted black line is the prior applied to $\Gamma$ during the spectral fitting step. All distributions are normalized to unit area.}
    \label{fig:gamma_parametric}
\end{figure}

The result of the reconstructed Gaussian parent distribution is shown in Fig.~\ref{fig:gamma_parametric}, as well as the inferred distributions using medians, stacked PDF and best-fit values from a maximum likelihood (ML) method (see Sect.~3.5 in Paper~\rom{1} for details). The parametric approach in general obtains a good fit with $\mu =1.82 \pm 0.02$, and $\sigma = 0.15 \pm 0.01$, except that its mean is about $3\sigma$ too high. The origin of this bias can be two-fold. First, there is a positive bias in the PDFs at low values of $\Gamma$, which is increasing when the true value deviates further from the value of 2.0 used as the median of the prior (see Fig.~10. in Paper~\rom{1}). This induces a bias in the fitted model as the parametric method cannot deal with biased PDFs. Second, the decrease of the prior of $\Gamma$ below 1.5 may bias the sources with very hard photon indices toward softer values, thereby shifting the mean of the fitted Gaussian. However, the effect is probably insufficient to explain the difference of more than $3\sigma$ with respect to the true value.

On the other hand, both the point estimator using the medians of the PDFs and the stacked PDF have trouble recovering the true distribution. The stacked PDF is too broad, giving too much probability for $\Gamma>2$. As already hinted above, the main reason is that each PDF is broadened by the uncertainty in the determination of $\Gamma$ \citep{Loredo2004}, especially for absorbed sources whose PDFs are hardly constrained (see Fig.~12. in Paper~\rom{1}) and close to the prior applied on $\Gamma$ during the spectral fitting process. On the other hand, the medians give a more constrained distribution compared to the stacked PDF. However, it is still much broader than the true one and, more importantly, it peaks in the bin $1.9\le\Gamma<2$, which is significantly higher than the true mean value (1.75), outside of the 3$\sigma$ range. The too high location of the peak is expected, because the median of $Gamma$ for objects with completely unconstrained posteriors (mostly absorbed) is 2. The ML fits tend to accumulate at the boundaries, as shown in Fig.~\ref{fig:gamma_parametric}, because in case of model degeneracy or poorly constrained parameters, the likelihood of a given parameter might be very flat, such that the ML solution may often be scattered outside of the allowed range (see Sects.\,4 and 5 of Paper~\rom{1} for more details). This shows that ML cannot be used to reconstruct the parent population. In this case, it is clear that the parametric fitting method is much better in recovering the parent distribution when we have loosely constrained PDFs.

\subsection{Forward-fitting population inference (FFPI)}
\label{new method}

\begin{figure}[tb]
    \centering
        \includegraphics[width=\columnwidth]{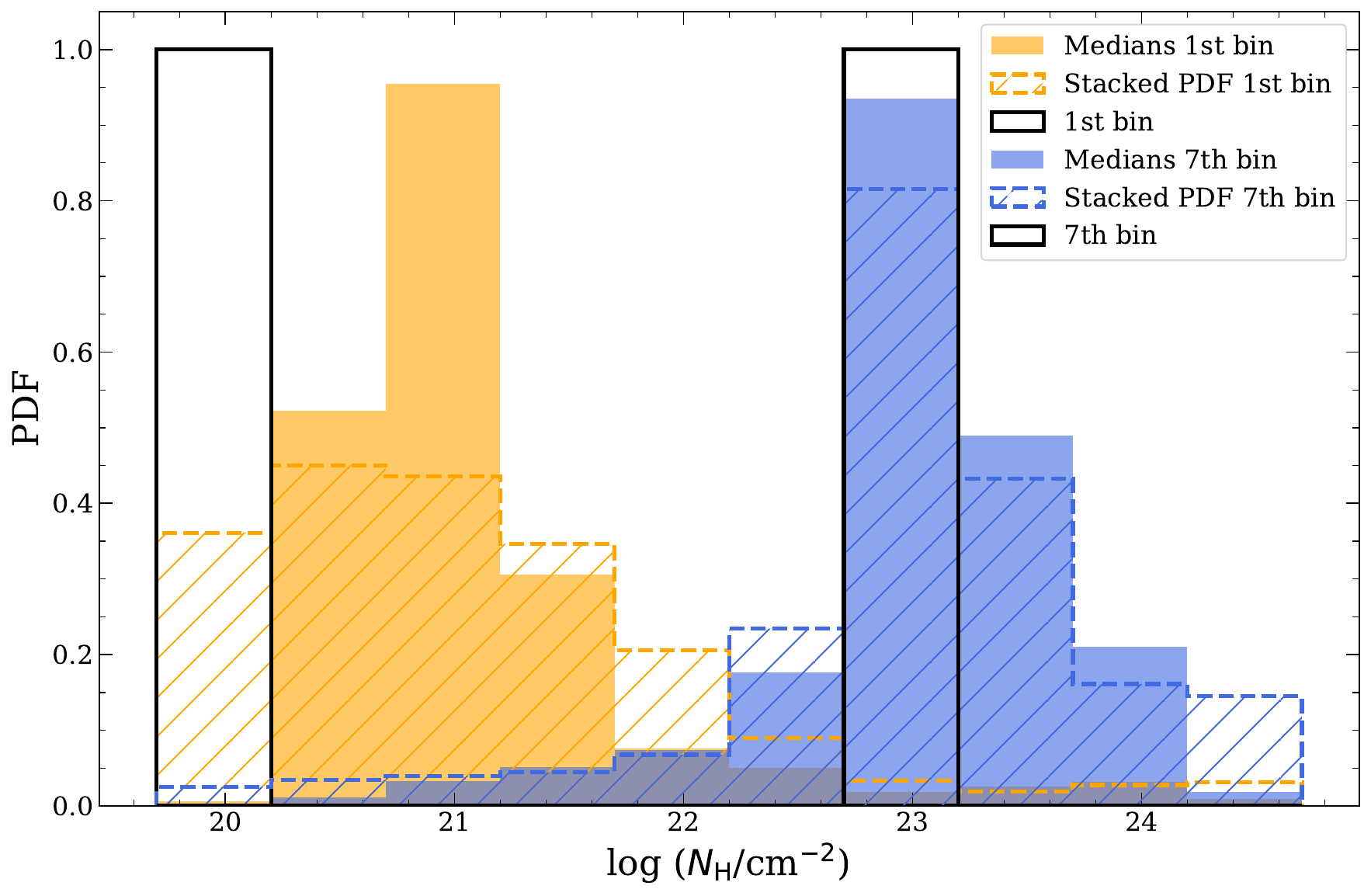}
    \caption{Stacked PDFs (hashed) and median distributions (filled) of the simulated sources with true $N_\mathrm{H}$ below $10^{20.2}$\,cm$^{-2}$ (first bin) shown in orange, and those with $N_\mathrm{H}$ from $10^{22.7}$ to $10^{23.2}$\,cm$^{-2}$ (seventh bin) shown in blue. The true $N_\mathrm{H}$ distributions of the two bins are shown in black (amplitude fixed to 1 for clarity). All distributions are normalized to unit area.}
    \label{fig:nh_redist_bin}
\end{figure}

The parametric method seemed to work well for the reconstruction of the parent distribution of $\Gamma$. However, it assumes a parametric form for the parent distribution, which may not be adapted to parameters like $N_\mathrm{H}$, for which no obvious parametric form exists. Moreover, the method implicitly assumes that the PDFs provide the true posterior probabilities of the parameters, namely that they are not biased, which is not the case for $N_\mathrm{H}$. As we discussed in Paper~\rom{1}, the PDF of $N_\mathrm{H}$ for an unabsorbed object is mostly flat below around $10^{22}$\,cm$^{-2}$, and steeply decreases above it, since the spectrum quickly becomes insensitive to this parameter in the unabsorbed regime. A similar effect can also be seen for highly absorbed sources. As a result, the PDFs of $N_\mathrm{H}$ of many sources are highly biased. For instance, the stacked PDF of sources in our simulated sample that have the true $N_\mathrm{H}$ below $10^{20.2}$\,cm$^{-2}$ is almost uniformly distributed below $10^{21.7}$\,cm$^{-2}$, and more importantly, the medians of the PDFs are completely off, peaking around $10^{21}$\,cm$^{-2}$, as shown in Fig.~\ref{fig:nh_redist_bin}. In this case the parametric method is not able to correct such a bias. On the other hand, both the stacked PDF and the medians of objects with $N_\mathrm{H}$ in the range of $10^{22.7}$ -- $10^{23.2}$\,cm$^{-2}$ are much more symmetric, and their peaks are located at the correct position. Therefore, the parametric method may reconstruct correctly the population properties for Compton-thin absorbed sources, but not for the less absorbed ones.

To overcome this difficulty we developed here a new method, the forward-fitting parameter inference (FFPI), to characterize the bias in the PDF. This method was inspired from the forward-fitting of X-ray spectra through a response, which is the standard method for analyzing X- and gamma-ray spectra with \texttt{xspec} \citep{Arnaud1996}, Sherpa \citep{Freeman2001}, SPEX \citep{Kaastra1996}, and so on. X- and gamma-ray spectral analysis faces indeed a very similar issue, as the recorded energies of mono-energetic photons present a tail at low energy, biasing the median recorded energy toward low values. Here, we forward-fit the stacked PDFs assuming a histogram model for the parent distribution.

We point out that Eddington bias is automatically taken into account in this forward-modeling approach, because the sources are scattered from their parent populations into the observed probability distributions using a fully representative propagation of errors.

\subsubsection{Transfer matrix}
\label{matrix}

\begin{figure}[tb]
\centering
\includegraphics[width=\columnwidth]{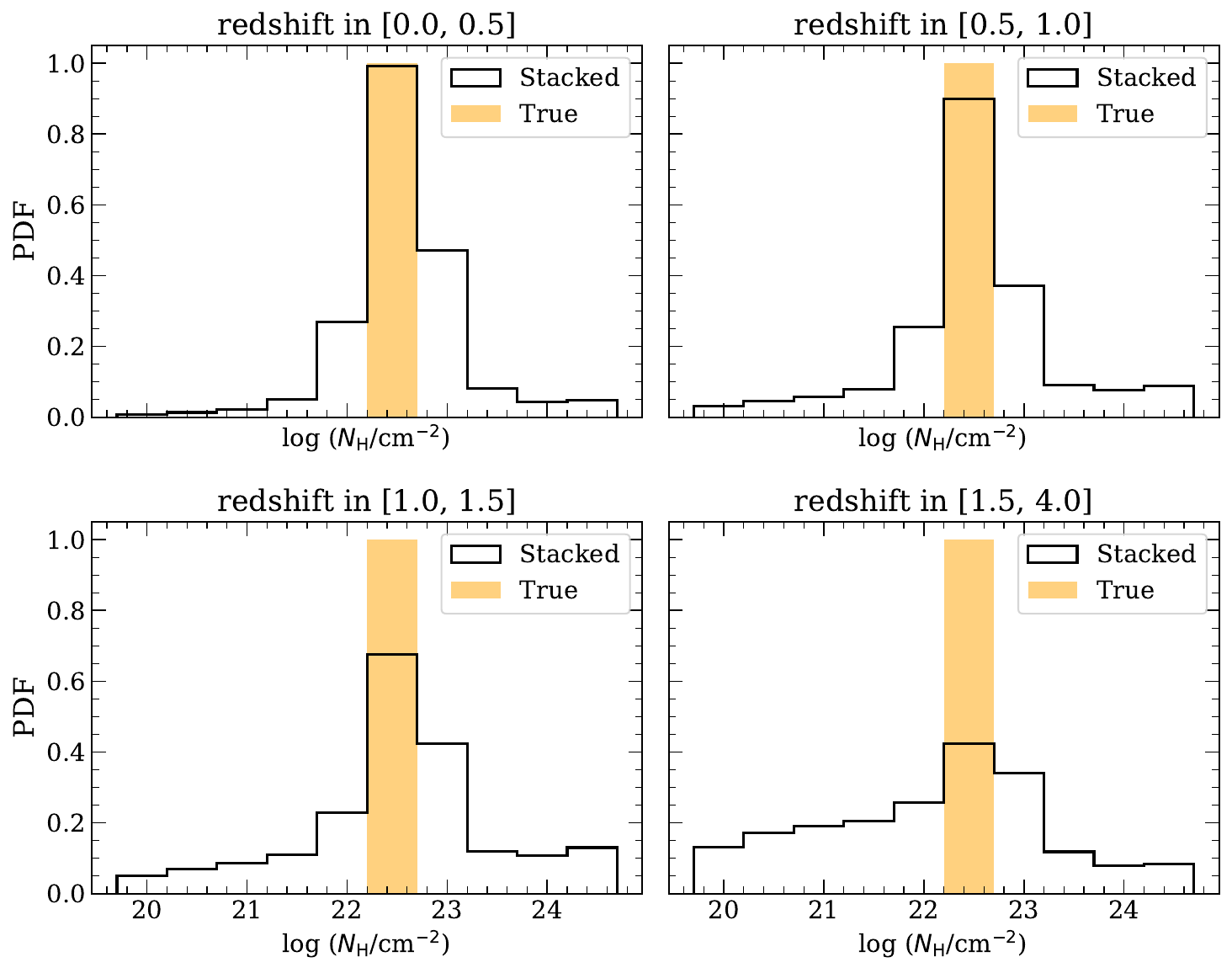}
\includegraphics[width=\columnwidth]{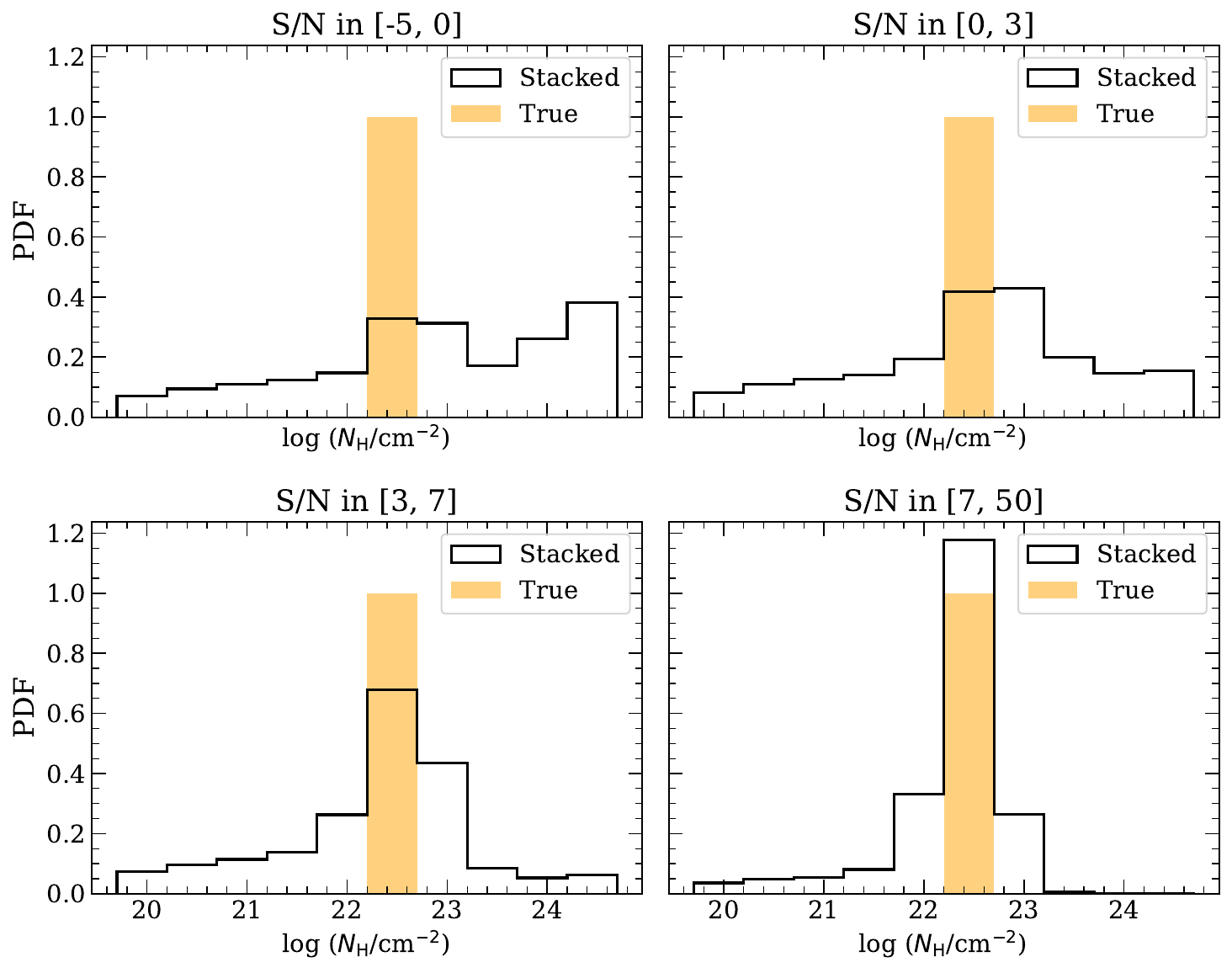}
\caption{Stacked PDFs of sources with true $N_\mathrm{H}$ in a single $N_\mathrm{H}$ bin in four different redshift subsamples (top four panels) and in four S/N subsamples (bottom four panels). The orange area shows the true $N_\mathrm{H}$ (amplitude fixed to 1 for clarity), and the black line shows the stacked PDF. We note that sources with negative S/N are formally detected by \texttt{ewavelet}, since the background is estimated differently. All distributions are normalized to unit area.}
\label{fig:nh_redist}
\end{figure}

The classic X-ray spectral fitting technique uses a transfer matrix called redistribution matrix file (RMF) that provides the probability of measuring a given energy as a function of the energy of an incident photon. The RMF can be determined from ground-based calibration experiments. Similarly, in our case we can determine the transfer matrix by computing the PDFs of simulated objects.

Nevertheless, the situation here is more complicated than in the case of the spectral analysis of X-ray photons, because the redistribution is not only dependent on the parameter itself, but also on the other spectral parameters. For example, the PDF of $\Gamma$ for a given object highly depends on $N_\mathrm{H}$. Theoretically, the transfer matrix should be computed for every possible set of parameters and provide the redistribution of all of them simultaneously. This would result however in a high-dimensional matrix that would be extremely expensive, first to compute and then to use in the fitting procedure. We can however greatly simplify the problem by marginalizing over the parameters that are not of immediate interest. For instance, if one wants to study the distribution of $\Gamma$, we can simulate a number of sources with fixed $\Gamma$, covering the whole range of interest, the other source parameters being chosen at random following prior distributions. Thus, this process remains essentially Bayesian, since it depends on the priors of the marginalized parameters.

The transfer matrix, $M_\mathrm{red}$, is defined in the following way. We represented the true parent population distribution of the parameter of interest as a one-dimensional histogram with an n-bin step function expressed by a vector $A_\mathrm{parent} = [p_1,p_2,\ldots,p_n]^{ \mathrm{T}}$, $\sum p_i =1$, where $^{ \mathrm{T}}$ is the transpose operator and $p_i$ represents the integrated probability in the $i^\mathrm{th}$ bin, and another vector for the stacked PDF $A_\mathrm{stacked} = [s_1,s_2,\ldots,s_n]^{\mathrm{T}}$, $\sum s_i =1$. Then $M_\mathrm{red}$ is defined as an $n \times n$ matrix such that
\begin{equation}
A_\mathrm{stacked} = M_\mathrm{red} \times A_\mathrm{parent}.
\end{equation}

Since this method uses matrix operations, a histogram model is required to model the parent distribution. However, a parametric parent distribution can be used, but it would need to be binned to create $A_\mathrm{parent}$. If we have a joint distribution of multiple parameters, the formalism can be also extended to higher dimensions, where $A_\mathrm{stacked}$ and $A_\mathrm{parent}$ would become $N-$dimensional arrays and $M_\mathrm{red}$ would have dimension $N+1$.

The parent distribution of $N_\mathrm{H}$ depends in particular on the redshift and the data quality, namely S/N. Since our spectral energy bands are fixed in the 0.6--12\,keV range, sources with different redshifts are observed in different energy bands in their local frame. This impacts the determination of $N_\mathrm{H}$, because photoelectric absorption is more efficient at low energies. When the spectrum is more and more redshifted, $N_\mathrm{H}$ is more and more difficult to constrain, except for very high absorption. This effect is shown in the top four panels of Fig.~\ref{fig:nh_redist}, taking as an example the stacked PDFs of $N_\mathrm{H}$ around $10^{22.5}$\,cm$^{-2}$ in four redshift bins. For sources with redshift smaller than 1, we clearly identify a peak around the true $N_\mathrm{H}$ value in the stacked PDF, but it decreases quickly as redshift increases, and more and more probability is redistributed to other bins. The S/N also strongly affects the PDFs, as shown in the bottom four panels of Fig.~\ref{fig:nh_redist}. The stacked PDF of sources with very high S/N (>7) is very well determined, with only a little leakage to lower $N_\mathrm{H}$ resulting from a few highly redshifted sources, while those with low S/N show a much less prominent peak at the true $N_\mathrm{H}$ and have clearly broader distributions. In particular, the stacked PDF of sources with negative S/N (although formally detected by \texttt{ewavelet}), where the background dominates, is almost flat across the range.

To construct $M_\mathrm{red}$, we used the simulated sample presented in Sect.~\ref{simulation}, which has been built so that its basic properties match those of the real \emph{XMM}-COSMOS data as well as possible; in particular, Figs.~\ref{fig:simul_lx_z} and \ref{fig:sample_cr_snr} show the excellent agreement between the redshift and S/N distributions.

\subsubsection{Forward-fitting}

Given $M_\mathrm{red}$ and the PDFs of the sources (and hence $A_\mathrm{stacked}$), the next step is to find the parent distribution $A_\mathrm{parent}$ through a forward-fitting procedure. First, we drew the parameters $p_{i}$ of $A_\mathrm{parent}$ from the prior distribution $P(A_\mathrm{parent})$. The prior we chose is an uninformative, flat histogram, that is to say,\ ${p_{i}}$ represents the coordinates of a simplex in $n$ dimensions ($n$ being the number of bins), whose vertices are the $n$ unit vectors. This can be achieved using a Dirichlet distribution $\mathrm{Dir}(\alpha_i)$, with $\alpha_{i}$ accounting for different sizes of the binning. In our case the binning is uniform, so $\alpha_{i}=1, \forall i$. Then we applied the transfer matrix, $M_\mathrm{red}$, to achieve the predicted distribution $A_\mathrm{red} = M_\mathrm{red} \times A_\mathrm{parent}$.

\begin{figure}[t]
    \centering
        \includegraphics[width=\columnwidth]{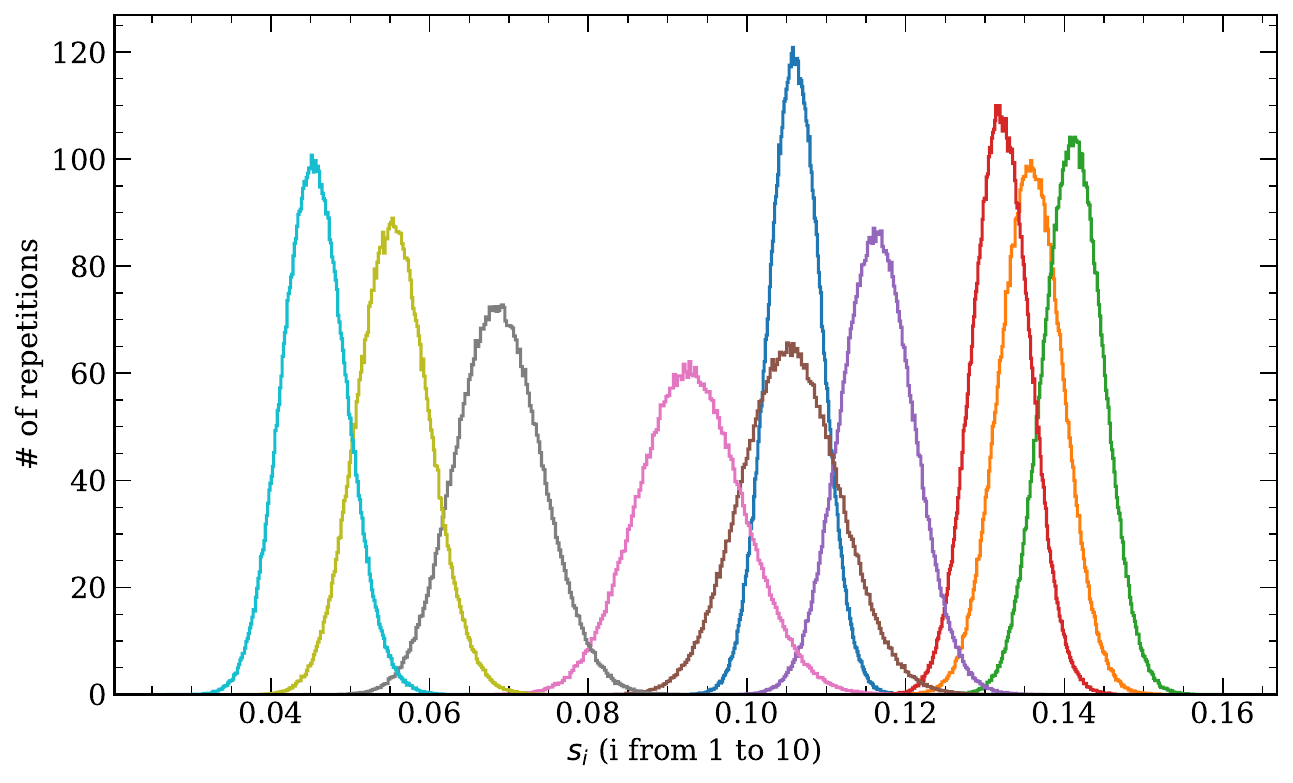}
    \caption{Distributions of the parameters $s_i$ of $A_\mathrm{stacked}$ in the bootstrapped data sample. The x axis shows the value of $s_i$, and the y axis shows the number of repetitions. The curves with different colors correspond to the different bins. All distributions are normalized to unit area.}
    \label{fig:bootstrap}
\end{figure}

Secondly, we need to determine the statistical properties of $A_\mathrm{stacked}$, so that we can compute the likelihood, which can be expressed as $P(A_\mathrm{stacked}|M_\mathrm{red},A_\mathrm{parent})$, which is equivalent to $P(A_\mathrm{stacked}|A_\mathrm{red})$. Since the theoretical distribution of the stacked PDF is unknown, we relied on the data sample itself to infer their statistics. We performed a bootstrap of the stacked PDF by drawing individual PDFs with replacement. We computed $10^6$ bootstrapped stacked PDFs, allowing us to construct the full statistical distribution of $A_\mathrm{stacked}$. An example of the distributions of $s_i$ of the ten-bin $N_\mathrm{H}$ stacked PDFs from a bootstrapped data sample is shown in Fig.~\ref{fig:bootstrap}. We find that the distributions of $s_i$ were reasonably well approximated with normal distributions. Thus, we can determine Gaussian uncertainties for all $s_i$ parameters, as well as their full covariance matrix using the bootstrapped sample. Finally, we sampled the posterior $P(A_\mathrm{stacked}|A_\mathrm{parent})$ with the No-U-Turn Markov chain Monte Carlo sampler (NUTS) implemented in the Python package \texttt{PyMC3} \citep{Salvatier2016}, because of its convenient implementation of the Dirichlet prior and faster sampling process compared to \texttt{MultiNest}, and we expected the chance of converging toward local maxima to be small. 

\subsubsection{Correction for selection effects}

The observed parameter distributions are affected by selection effects. The true distribution of observed parameters indeed differs in general from that of the full population; for instance, absorbed objects are more difficult to detect than unabsorbed ones. The parameter distribution of the full population can be reconstructed from that obtained with FFPI by determining, for each bin of the distribution, the fraction of simulated objects that are detected. Since we know, for each simulated source, whether it has been detected or not, we can do this by directly counting the detected objects in the bin, without using the selection function from Eq.~(\ref{eq:selfuc}) explicitly.

We note that this procedure will amplify the counting noise in the bins where the selection effect is very strong. This is clearly the case, for instance, for bins of the column density distribution at very large values of $N_\mathrm{H}$.

\subsection{Validation of FFPI}
\label{Validation of FFPI}

\subsubsection{Hydrogen column density $N_\mathrm{H}$}

\begin{figure}[tb]
\centering
\includegraphics[width=\columnwidth]{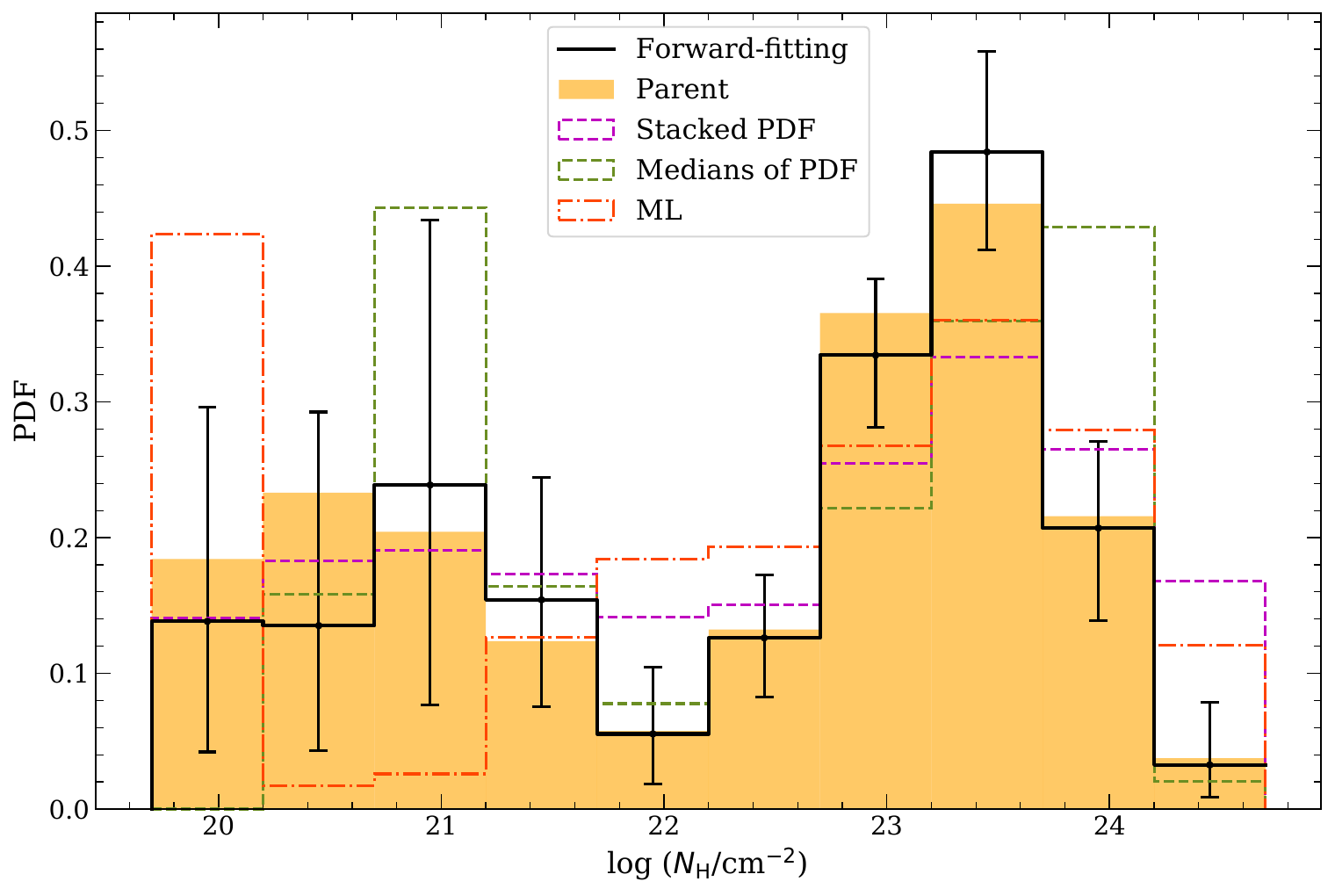}
\includegraphics[width=\columnwidth]{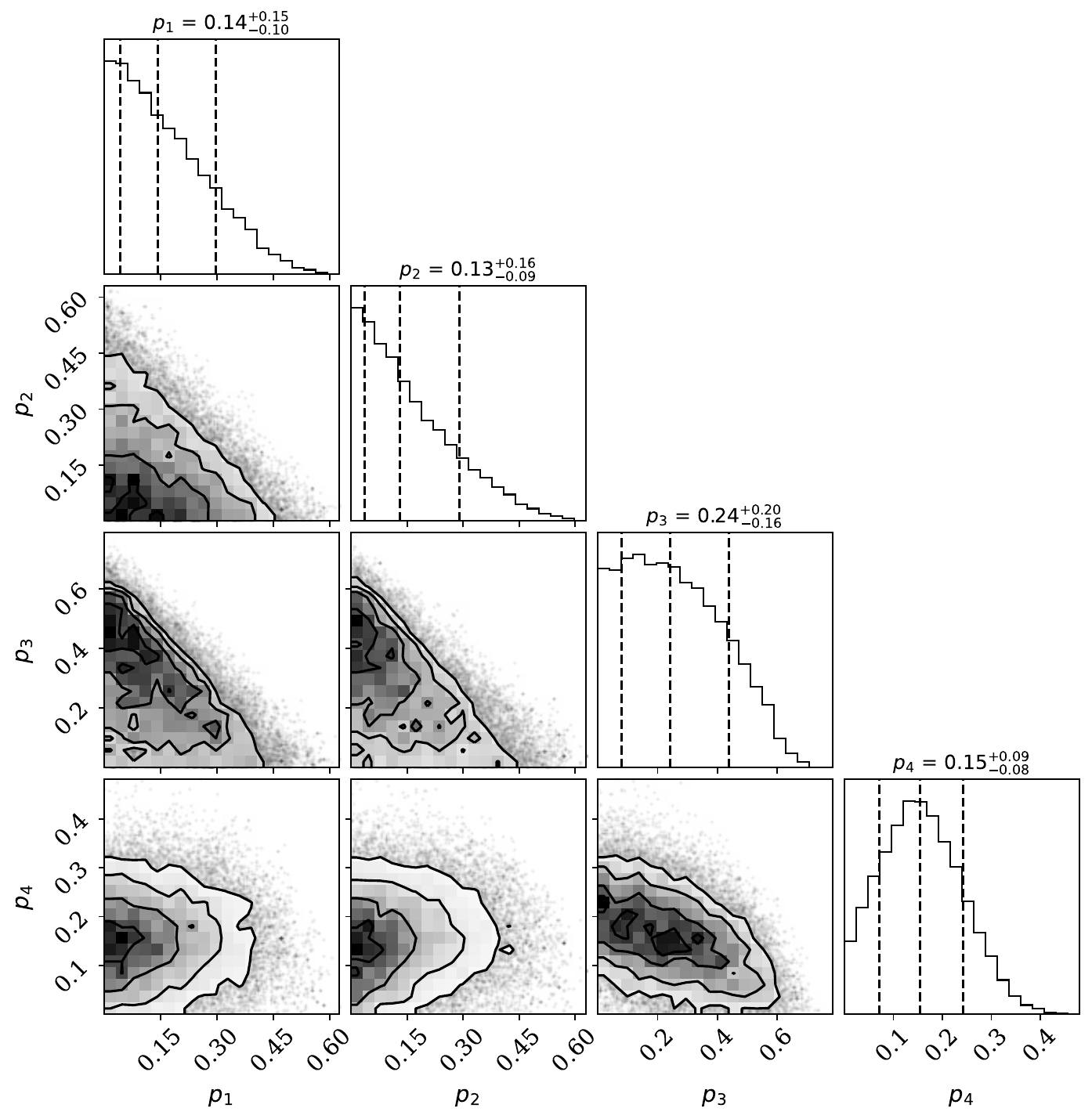}
\caption{Parent distributions of $N_\mathrm{H}$. Upper panel: Parent distributions for a subsample that is the same size as the \emph{XMM}-COSMOS data sample and is randomly selected from the simulation. The true distribution is plotted as the filled histogram in orange, and the reconstructed $N_\mathrm{H}$ distribution obtained by applying FFPI is shown as the black line. The error bars show the 68.3\% credible intervals. For comparison, the stacked PDFs, the distributions of medians, and the ML results are plotted in dashed magenta, dashed green, and dash-dotted red lines, respectively. All distributions are normalized to unit area. Lower panel: Corner plot of the parameters of the first four bins ($p_1$ -- $p_4$, converted to probability density for consistency with the lower panel) in the unabsorbed regime.}
\label{fig:simul_nh_redist}
\end{figure}

We first tested this method on $N_\mathrm{H}$ by constructing a subsample of the simulation with the same number of sources as the \emph{XMM}-COSMOS data sample, keeping the same redshift and S/N distributions, so that the same transfer matrix can be used, which is only one-dimensional in this case. However, we made its true $N_\mathrm{H}$ distribution follow a double Gaussian distribution, as shown in the lower panel of Fig.~\ref{fig:simul_nh_redist}. The reason for choosing this distribution is two-fold. First, with such complex distribution we can clearly see the performance of the method on absorbed and unabsorbed sources; and second, it is suggested that there may be a similar bimodality in the $N_\mathrm{H}$ parent distribution of real sources \citep{Paltani2008,Ueda2014,Ricci2017}.

The inferred parent distribution by FFPI is shown in the same figure, together with the results obtained using medians of the PDFs, the stacked PDFs and ML. In the absorbed regime, our approach successfully reconstructed the parent distribution of $N_\mathrm{H}$  with fairly small uncertainties, while the other methods either have the peak at a wrong place (medians) or severely underestimate it due to the broadening of the distribution. The bin of Compton-thick sources (corresponding to sources with $N_\mathrm{H}$ above $1.6 \cdot 10^{24}$\,cm$^{-2}$) and the valley around $10^{22}$\,cm$^{-2}$ are also very well constrained, clearly recovering the bimodality. In the unabsorbed regime, the uncertainties become much larger and strongly correlated, as shown in the corner plot \citep{corner} of Fig.~\ref{fig:simul_nh_redist}. The first three bins ($p_1$ -- $p_3$) are basically recovering the Dirichlet prior, while $p_4$ starts to be better constrained and decoupled from $p_1$ and $p_2$. However, despite the little information that we obtain for unabsorbed sources, the result correctly indicates that it is quite difficult to constrain the distribution of $N_\mathrm{H}$ below $10^{22}$\,cm$^{-2}$. By contrast, the distribution of medians gives us a false peak at $N_\mathrm{H}\sim 10^{21}$\,cm$^{-2}$, while the stacked PDF is not able to distinguish the bimodality, since it is simply flat until almost $ 10^{23}$\,cm$^{-2}$ and has a much lower peak. The ML results are even worse, with a very high peak in the lowest bin due to the fact that almost all unabsorbed sources are fitted with $N_\mathrm{H}$ at the lower boundary.

\subsubsection{Photon index $\Gamma$}

To reconstruct the distribution of the photon index $\Gamma$ with FFPI, we need to extend the transfer matrix to two parameters, because the measurement of $\Gamma$ strongly depends on $N_\mathrm{H}$, as shown in figure~11 of Paper~\rom{1}, and the flat $N_\mathrm{H}$ distribution assumed in the simulation is probably very different from the reality. While in principle using more bins could be more accurate, here we simply used an $8\times2$ two-dimensional histogram model for the joint distribution of $\Gamma$ and $N_\mathrm{H}$, because our data sample has a limited number of sources and our main goal for this exercise is to reconstruct the distribution of $\Gamma$. The two bins in the $N_\mathrm{H}$ dimension were separated at the boundary between absorbed and unabsorbed sources, at $N_\mathrm{H}=10^{22}$\,cm$^{-2}$, while for $\Gamma$ the bins were uniform from 1.2 to 2.6.

\begin{figure}[tb]
\centering
\includegraphics[width=\columnwidth]{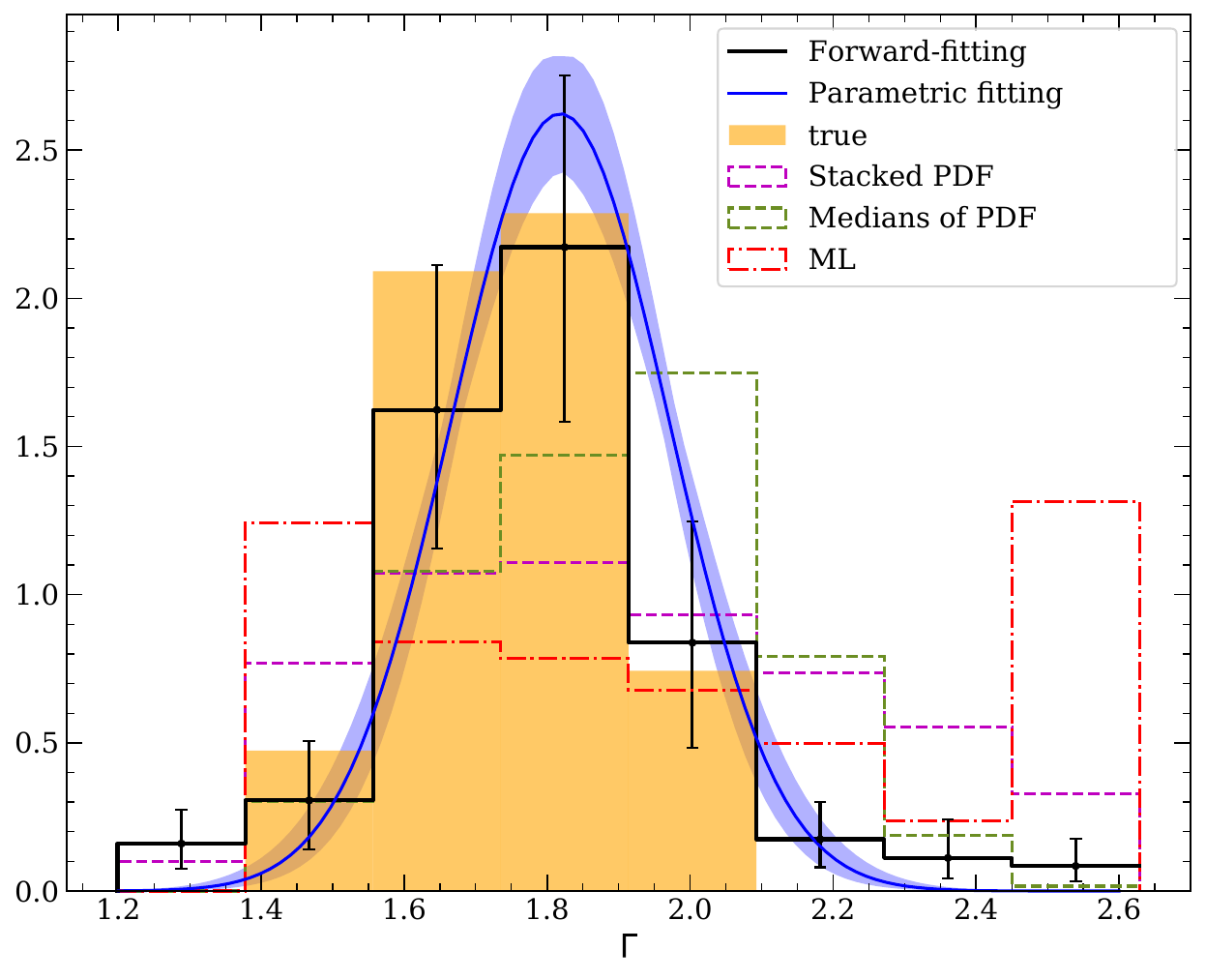}
\includegraphics[width=\columnwidth]{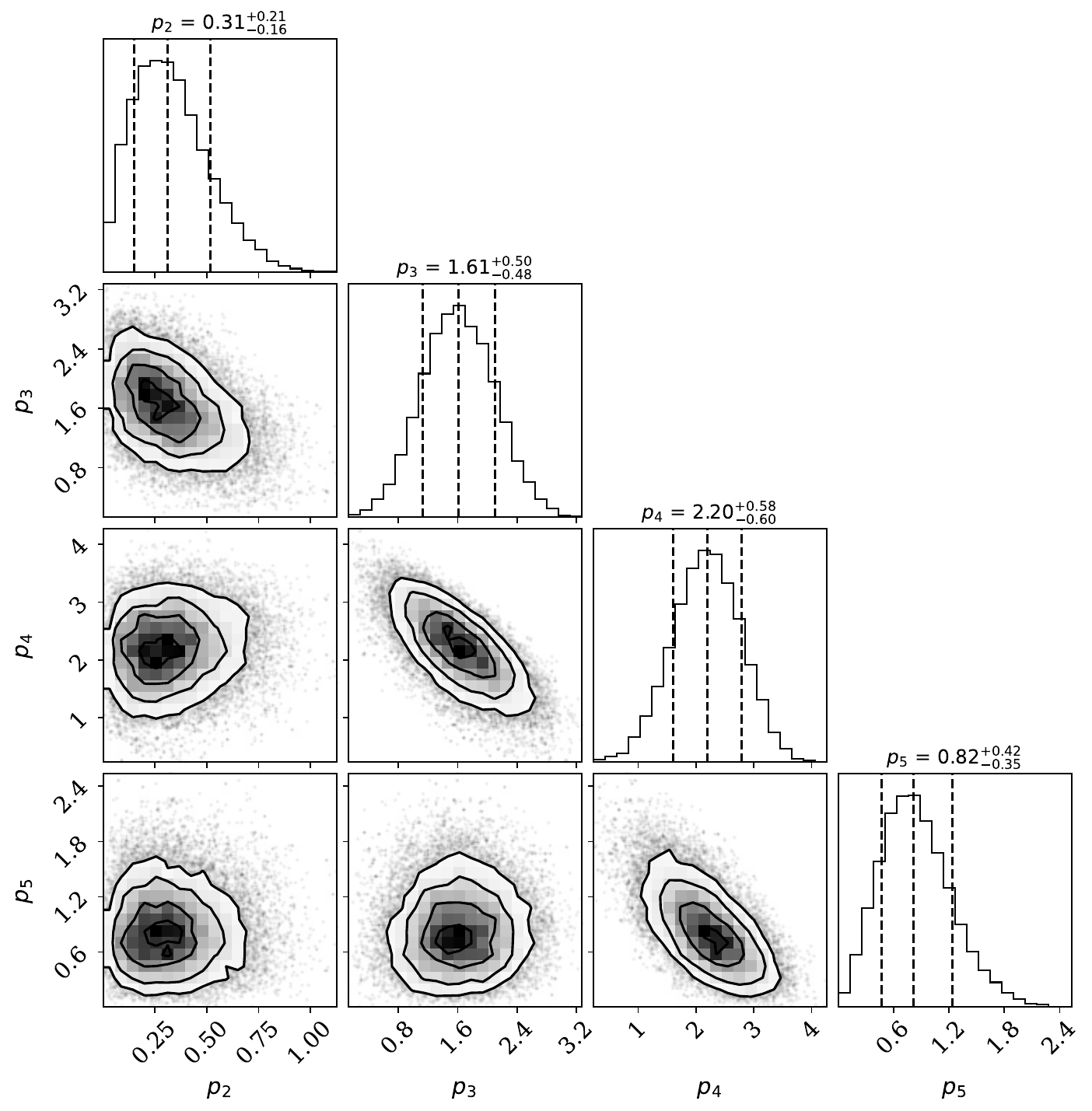}
\caption{Parent distributions of $\Gamma$. Upper panel: Parent distributions for a subsample that is the same size as the \emph{XMM}-COSMOS data sample and is randomly selected from the simulation. The true distribution is plotted as the filled histogram in orange, and the reconstructed $\Gamma$ distribution obtained by applying FFPI is shown as the black line. We also show the reconstructed distribution of $\Gamma$ using the Bayesian parametric fitting in blue. Error bars and the shaded blue area show the 68.3\% credible intervals. All distributions are normalized to unit area. Lower panel: Corner plot of the parameters of the second to the fifth bins ($p_2$ -- $p_5$, converted to probability density for consistency with the lower panel), which include the vast majority of the sources.}
\label{fig:simul_gamma_recon}
\end{figure}

\begin{figure*}[tb]
\centering
\includegraphics[width=17cm]{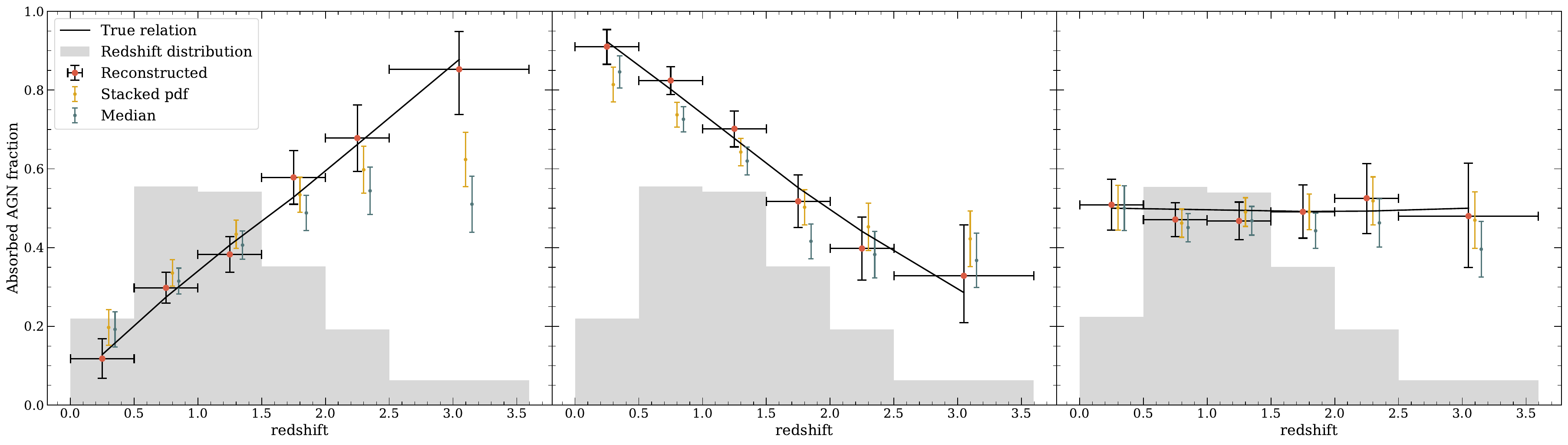}
\includegraphics[width=17cm]{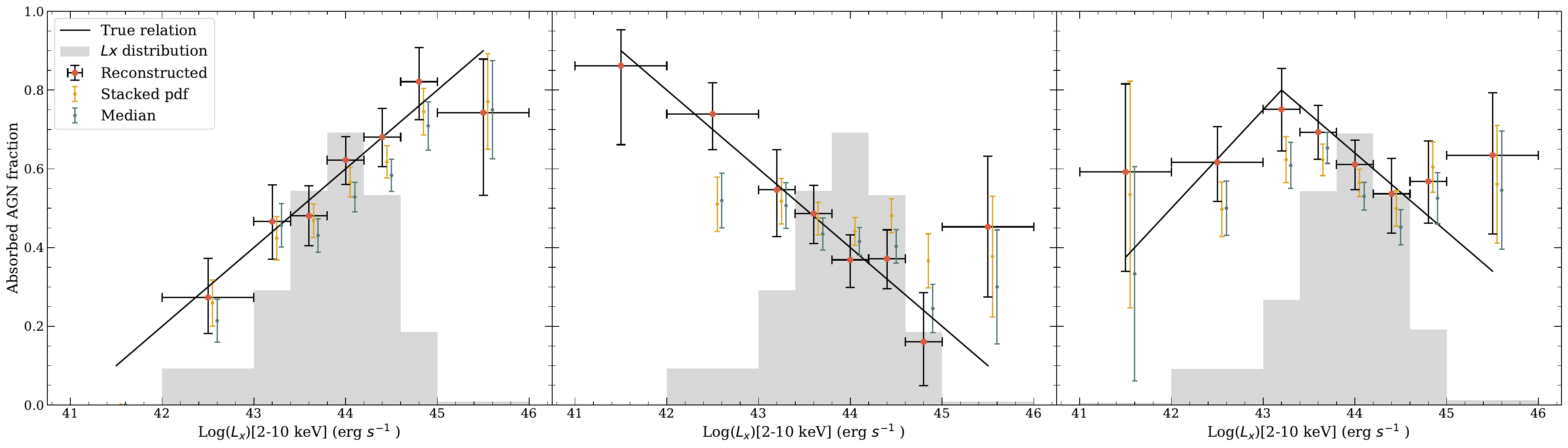}
\caption{Reconstruction of the absorbed AGN fraction versus redshift (top row) and $L_\mathrm{X}$ (bottom row) for different subsamples of the simulation. The black line represents the true absorbed fraction relation in each redshift or $L_\mathrm{X}$ bin, and the reconstructed one obtained by applying our method is plotted with red dots and vertical error bars showing the 68.3\% credible intervals. The results of using stacked PDF and medians are shown in yellow and green, respectively. The gray histogram represents the $L_\mathrm{X}$ or redshift distributions of the simulated sample. }
\label{fig:obsfrac_simul}
\end{figure*}

We tested FFPI on the same sample as introduced in Sect.~\ref{parametric method}, and the result is illustrated in Fig.~\ref{fig:simul_gamma_recon}, together with the one using the parametric method. The true $\Gamma$ distribution is correctly recovered by FFPI and the result in general agrees with that of the parametric one. However, our method corrects for biased PDFs whereas the parametric method does not. This allows FFPI to recover the peak more precisely, as the median of the distribution reconstructed by FFPI is around the true value (1.75), whereas the parametric method biases $\Gamma$ toward larger values, as discussed in Sect.~\ref{parametric method}.

On the other hand, the uncertainties on the histogram distribution of FFPI are larger than those on the Gaussian distribution, especially for the two bins around the peak, but this is expected, because FFPI has many more free parameters ($8\times2$ versus 2), since the parent distribution is nonparametric and specified with an arbitrary histogram. Also, the assumption of the Gaussian parent distribution is a much stronger prior than the non-informative one on the histogram model. It works well when the parent distribution is indeed a Gaussian, like this test sample, but with FFPI we keep the possibility of finding alternative distributions. Actually, in FFPI we have the freedom to adjust the informative level of the prior by changing the concentration parameter $\alpha$ of the Dirichlet distribution. For example, if we believe that very few objects would have $\Gamma$ close to the boundaries of the parameter space, we can increase $\alpha$ to give more weight to the bins around the center and penalize those close to the boundaries, imposing an approximate Gaussian prior on the parent distribution of $\Gamma$.

The corner plot \citep{corner} in Fig.~\ref{fig:simul_gamma_recon} shows the PDFs of $p_2$ through $p_5$, which contain all the sources, and their correlations. The adjacent bins are strongly anticorrelated, especially the two bins around the peak, which is a natural outcome of applying the transfer matrix. On the other hand, bins that are not neighbors are clearly uncorrelated. This shows that the matrix of $\Gamma$ is not far from being diagonal, and mainly redistributes the probability into the neighboring bins.

\subsubsection{Absorbed fraction versus $z$ and X-ray luminosity $L_\mathrm{X}$ }

We also test the performance of FFPI to reconstruct the evolution of the absorbed fraction with redshift, as well as its dependence on the 2--10\,keV intrinsic X-ray luminosity ($L_\mathrm{X}$). The absorbed fraction is defined as the fraction of sources whose spectral parameter $N_\mathrm{H}$ is larger than $10^{22}$\,cm$^{-2}$. In this specific case we focus only on the classification between absorbed and unabsorbed sources. Therefore, similarly to $\Gamma$, the transfer matrix is two-dimensional, with only two bins in the $N_\mathrm{H}$ space with boundary at $10^{22}$\,cm$^{-2}$. The transfer matrix needs however to be constructed for several bins in redshift and luminosity, respectively.

To infer the redshift evolution of the absorbed fraction, we divided the total sample into different redshift bins and then reconstructed the $N_\mathrm{H}$ distribution modeled by the two-bin histogram model in each redshift bin. The transfer matrices of different redshift bins were also derived from the total simulated sample divided into the same bins. In order to validate this approach, we constructed three subsamples from the simulation; all of them share similar redshift and S/N distributions to those of our \emph{XMM}-COSMOS data sample, as well as the same sample size. We imposed an increasing absorbed AGN fraction with redshift in the first sample, while the other two have a decreasing and a flat relation, respectively. We made six bins in the redshift space, with the first five bins being equally spaced with a bin width of 0.5, and the last bin having a width of 1.1 to encompass the most distant source ($z=3.6$). This binning was chosen to optimize the resolution in redshift, as well as the number of sources in each bin, for our \emph{XMM}-COSMOS sample. The results, as well as those using medians or stacked PDFs, are shown in the first row of Fig.~\ref{fig:obsfrac_simul}. For the medians it was straightforward to calculate the absorbed fraction, while for the stacked PDF, we used the stacked posteriors integrated above $10^{22}$\,cm$^{-2}$ as an estimate for the absorbed fraction. We see that, using medians or stacked PDFs, we tend to underestimate the slope of the relation. This is particularly clear in the case of an increasing absorbed fraction with redshift, where at high redshift the stacked PDF and medians substantially underestimate the absorbed fraction (upper left panel of Fig.~\ref{fig:obsfrac_simul}), due to the increasing difficulty of constraining $N_\mathrm{H}$ as redshift increases (see Sect.~5.3 in Paper~\rom{1}). On the other hand, our approach successfully recovers the true absorbed fraction without measurable bias for all three samples.

Regarding the absorbed fraction versus $L_\mathrm{X}$, we had to reconstruct the joint $L_\mathrm{X}$--$N_\mathrm{H}$ distribution, since the intrinsic $L_\mathrm{X}$, which depends on most of the spectral parameters and on the redshift, is unknown. Although the spectral method recovers relatively well the true $L_\mathrm{X}$ (see Paper~\rom{1} for details), there could still be a biased relation because of the bias in the classification of absorbed sources. Therefore, a transfer matrix that is able to take care not only of the bias of $N_\mathrm{H}$, but also of $L_\mathrm{X}$, is required. The same two-bin model of $N_\mathrm{H}$ was applied to derive the absorbed fraction, and we used a non-uniform 8-bin model for the $L_\mathrm{X}$ distribution, with broad bins at very low and very high $L_\mathrm{X}$ (two bins below $10^{43}$\,erg\,s$^{-1}$ and one bin above $10^{45}$\,erg\,s$^{-1}$), where few sources are detected, and finer bins in between (5 equal-width bins).

We also tested this method using three subsamples from the simulation, as shown in Fig.~\ref{fig:obsfrac_simul}, which have different absorbed fraction versus $L_\mathrm{X}$ relations, but the same sample size and total $L_\mathrm{X}$ distribution as the \emph{XMM}-COSMOS data sample. Specifically, we imposed a positive relation in the first subsample, a negative one in the second, and the last one having a peak of absorbed fraction at luminosity $10^{43.3}$\,erg\,s$^{-1}$. The results show that, for the first subsample, the inferred fractions using the stacked PDF or the medians are always biased low, while for the second one, the slope of the relation is clearly biased toward flatter values. In the third case, the two approaches also bias the fractions low, but the impact is stronger than in the previous two cases, as the total relation becomes simply a flat one. In particular, the recovered value around the peak at $\log L_X=43$ is more than 2$\sigma$ away from the true value. By contrast, FFPI is able to reconstruct the absorbed fraction more accurately with reasonable uncertainties in the bins with $L_\mathrm{X}$ between $10^{42}$ to $10^{45}$\,erg\,s$^{-1}$ for all three subsamples, and in the third situation, there is a clear hint for a peak of the absorbed fraction at the right position, despite the large errors in the two extreme bins due to the scarcity of sources in the corresponding bins.

\section{Analysis of the \emph{XMM}-COSMOS sample}
\label{results}

The Bayesian spectral fitting method introduced in Paper~\rom{1} was applied to the 819 sources in our COSMOS data sample (see Sect.~\ref{cosmos}) to obtain the PDFs of the spectral parameters of each source. We reconstructed here the parent distributions of $N_\mathrm{H}$ and $\Gamma$, as well as the absorbed fraction versus 2--10\,keV $L_\mathrm{X}$ and redshift, using the methods described in Sect.~\ref{method}. Unless specified otherwise, the results are corrected for the selection effects determined based on the simulations (see Sect.~\ref{simulation}).

\begin{figure*}[htb]
    \centering
        \includegraphics[width=17cm]{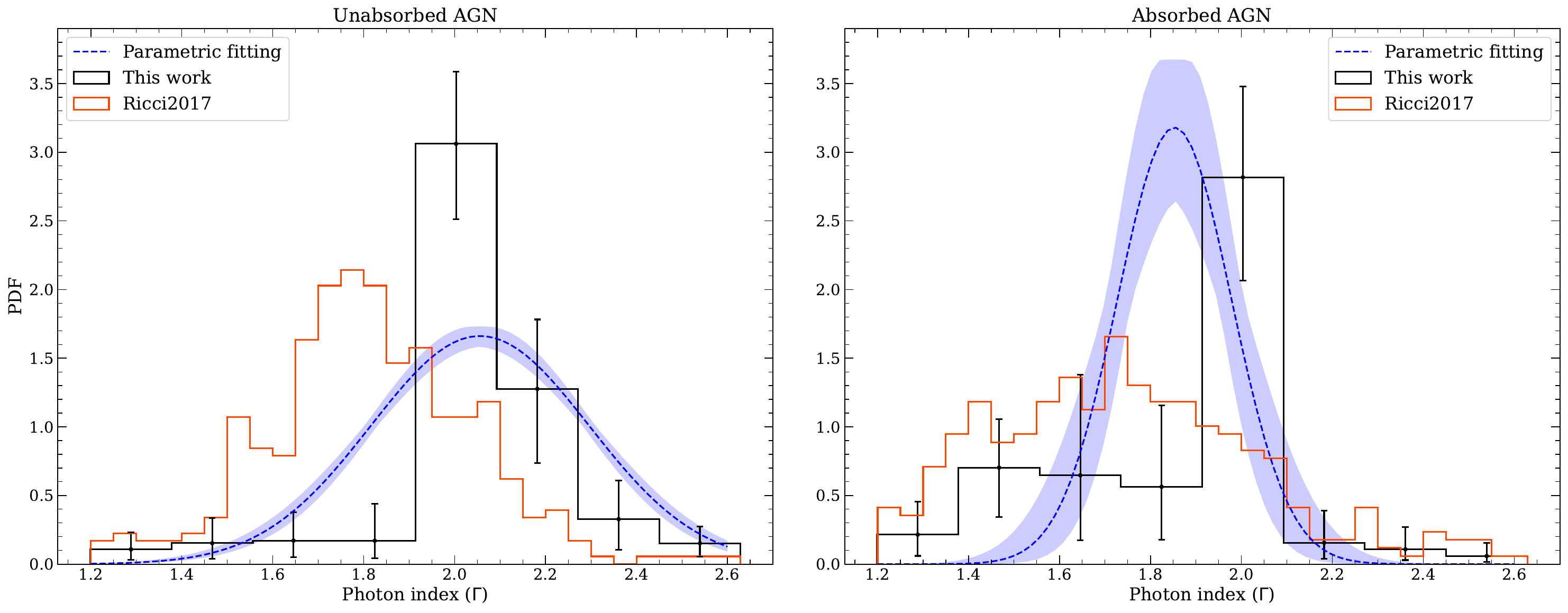}
    \caption{Inferred parent distributions of the photon index, $\Gamma$, for unabsorbed AGN (left) and absorbed AGN (right). Our results are shown in black, with error bars representing 68.3\% credible intervals. The results of \citet{Ricci2017} are plotted in red, and the dashed blue lines show those from the parametric fitting, with the filled area also showing 68.3\% credible intervals. All distributions are normalized to unit area.}
    \label{fig:gamma_parentfit}
\end{figure*}

\begin{figure}[tb]
    \centering
        \includegraphics[width=\columnwidth]{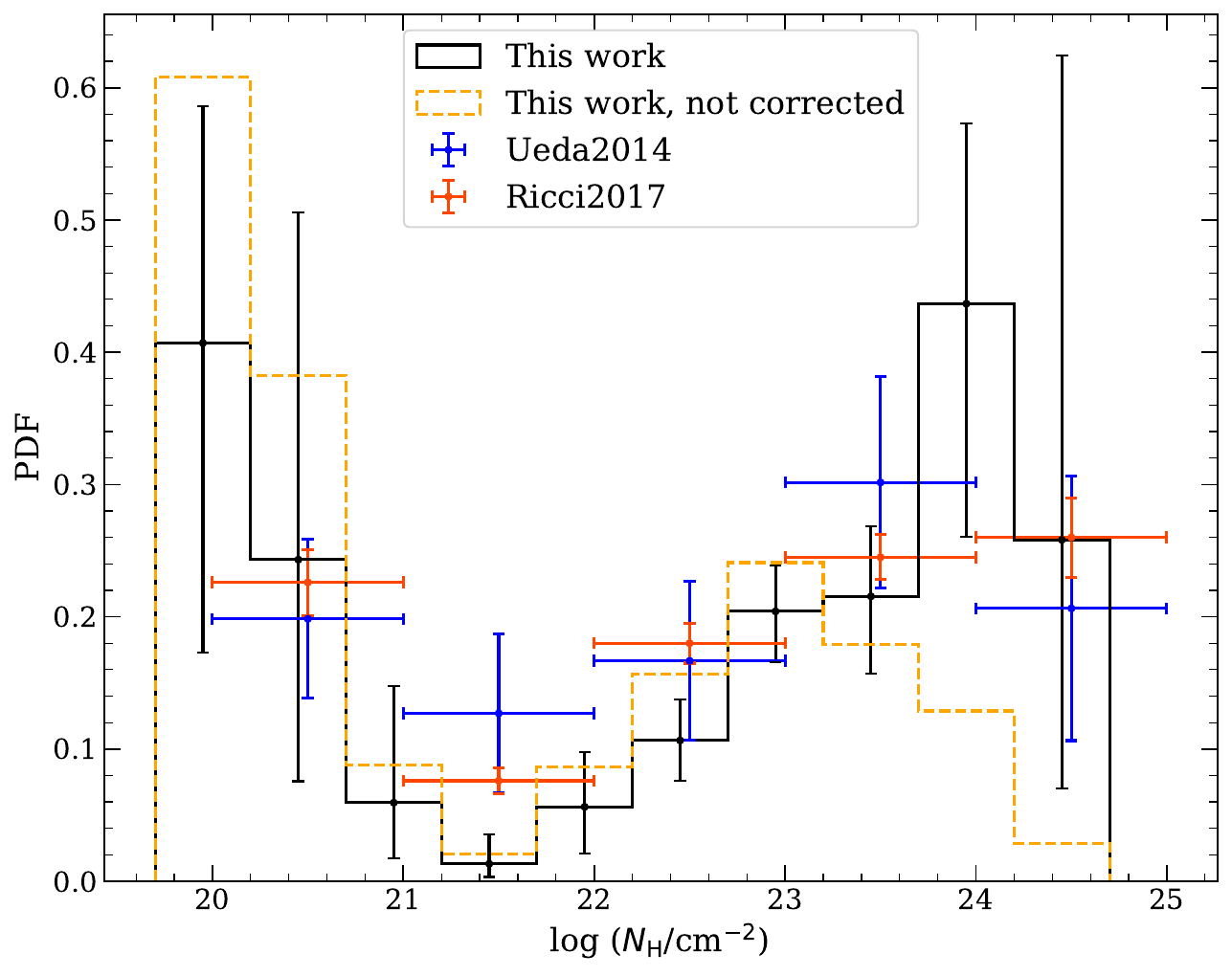}
    \caption{Observed distribution of $N_\mathrm{H}$ inferred with our method (dashed orange line), and corrected for selection effects (solid black line), with vertical error bars representing 68.3\% credible intervals. The blue and red points are the results from \citet{Ueda2014} and \citet{Ricci2017}, respectively. The horizontal error bars show the bins. All distributions are normalized to unit area.}
    \label{fig:nh_parentfit}
\end{figure}

\begin{figure}[tb]
    \centering
        \includegraphics[width=\columnwidth]{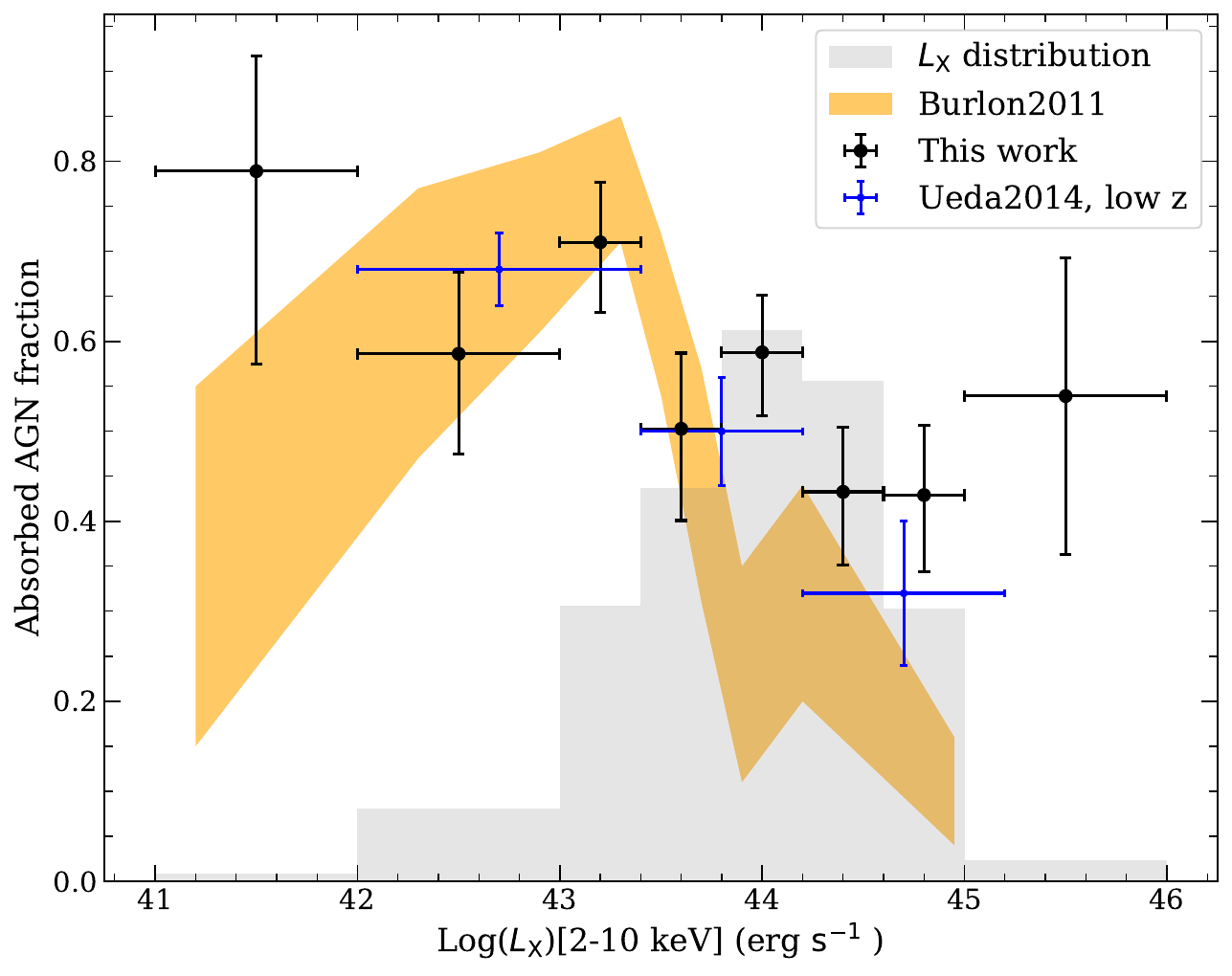}
    \caption{Absorbed fraction versus 2--10\,keV luminosity. Black points are the result of this work, taking selection effects into account; the vertical error bars show 68.3\% credible intervals, and the horizontal error bars show the binning. The blue points are the results of \citet{Ueda2014} for low-redshift sources, and the orange area shows the results of \citet{Burlon2011}. The gray area is the observed luminosity distribution in our sample.}
    \label{fig:obsfrac_lx}
\end{figure}

\begin{figure}[tb]
    \centering
        \includegraphics[width=\columnwidth]{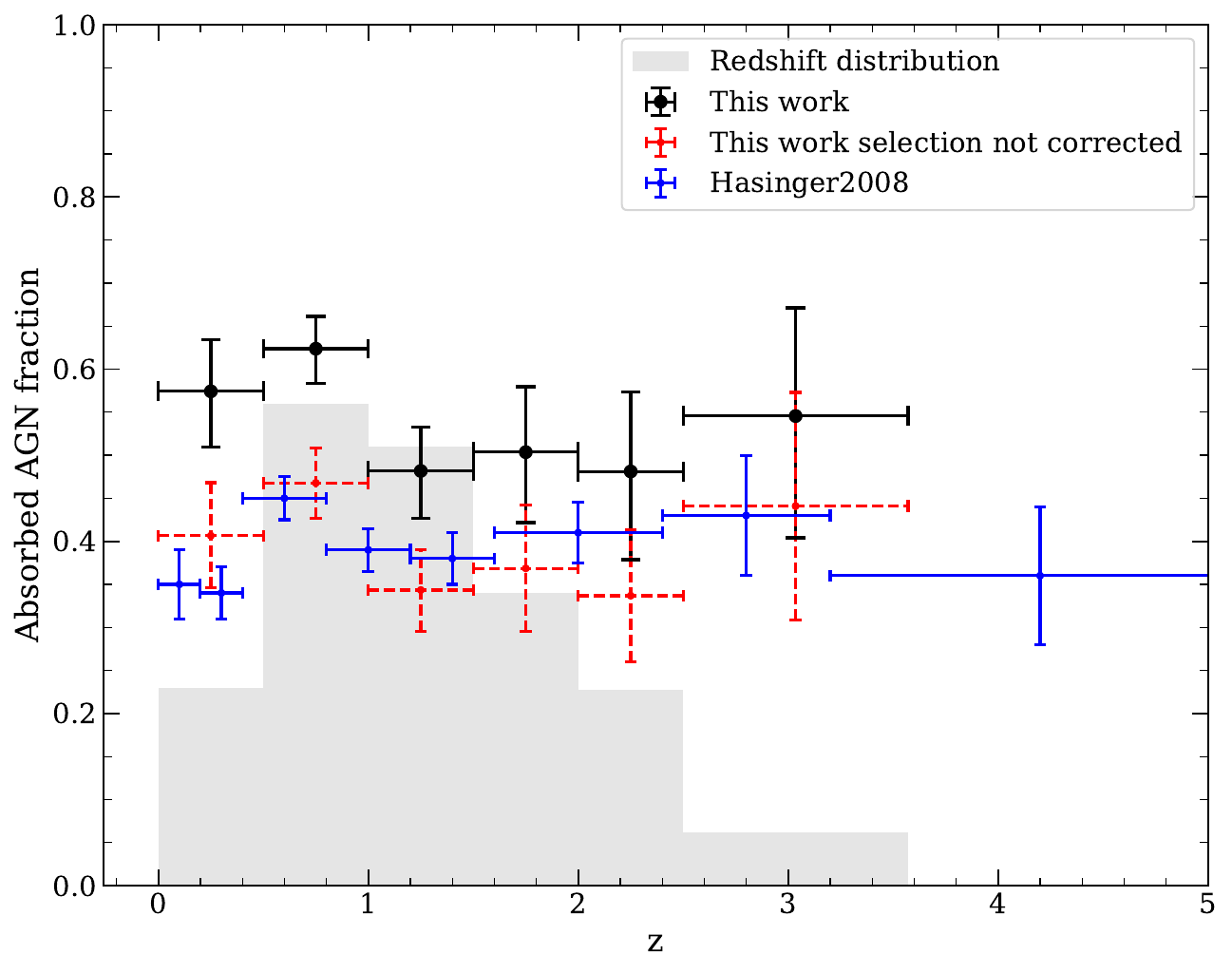}
    \caption{Evolution of the absorbed fraction with redshift. Our results corrected for selection effects are shown in black. The results without selection correction are shown in red with dashed error bars. Vertical error bars show 68.3\% credible intervals; horizontal error bars show the binning. The results of \citet{Hasinger2007} are shown in blue. The gray area shows the redshift distribution of our sample, including both spec-z and photo-z.}
    \label{fig:obsfrac_z}
\end{figure}

The inferred distributions of photon indices $\Gamma$ for absorbed and unabsorbed AGN using FFPI are shown in Fig.~\ref{fig:gamma_parentfit}, as well as those using the parametric approach. With FFPI we find relatively non-Gaussian distributions for both the absorbed and unabsorbed AGN in our sample, with a dominant peak in the bin around $\Gamma=2$, containing around 50\% of the total probability, but the rest is distributed  differently. The unabsorbed population has an excess of probability to have a source at $\Gamma$ higher than the peak, but very low probability below it, while the opposite is true for the absorbed population. The selection effects on $\Gamma$ are quite small, so that the corrected and uncorrected distributions are very similar; thus only the corrected distributions are shown in Fig.~\ref{fig:gamma_parentfit}. On the other hand, the standard approach obtains a much broader Gaussian distribution for the unabsorbed population with $\mu=2.05\pm0.01$ and $\sigma=0.24\pm0.01$, and a relatively narrow one for the absorbed population, with $\mu=1.85\pm0.03$ and $\sigma=0.12\pm0.02$. It also obtains in general a harder distribution of $\Gamma$ for absorbed AGN, which is consistent with the results of FFPI.

The reconstructed $N_\mathrm{H}$ distribution is shown in Fig.~\ref{fig:nh_parentfit}. We can see a clear bi-modality with two well separated peaks, one at $10^{20}$\,cm$^{-2}$ and the other at $10^{24}$\,cm$^{-2}$. Most of the unabsorbed sources have $N_\mathrm{H}$ below $10^{20.7}$\,cm$^{-2}$, with a tentative decreasing trend from $10^{20.2}$ to $10^{21.7}$\,cm$^{-2}$. Between $10^{20.7}$ and $10^{22.2}$\,cm$^{-2}$, there are very few sources and the distribution is compatible with zero, while it increases steadily after entering the absorbed regime. The uncertainties are relatively large for bins at both ends of the distribution. For the unabsorbed ones the uncertainties are mainly the result from unconstrained PDFs, as discussed in Sect.~\ref{method}, while for the most absorbed ones the uncertainties are due to the increasing effect of the selection correction as absorption gets stronger. The bin with the least probability to have a source is around $10^{21.5}$\,cm$^{-2}$, which is slightly below the adopted boundary between absorbed and unabsorbed regimes, conventionally set at $10^{22}$\,cm$^{-2}$, but agrees with that proposed by \citet{Merloni2014}.

The reconstructed relation between the absorbed AGN fraction and the intrinsic X-ray luminosity in the 2--10\,keV band is illustrated in Fig.~\ref{fig:obsfrac_lx}. We find the well-known anticorrelation between the two in the range of $10^{43}$ -- $10^{45}$\,erg\,s$^{-1}$. The tentative peak is at $10^{43.2}$\,erg\,s$^{-1}$, where the absorbed fraction is around 0.7. The highest luminosity bin ($10^{45}$ -- $10^{46}$\,erg\,s$^{-1}$) shows a small increase, compatible with no change in the absorbed fraction considering its large uncertainty. This can be a statistical fluctuation due to the paucity of very luminous sources in our sample, as shown by the gray area in Fig.~\ref{fig:obsfrac_lx}. For very faint AGN that are less luminous than $10^{43.2}$\,erg\,s$^{-1}$, the behavior is very uncertain because of very few sources and large selection correction.

The inferred absorbed fraction as a function of redshift is shown in Fig.~\ref{fig:obsfrac_z}. The redshift of our data is assumed to be known because our sources have either spec-z or good photo-z \citep{Salvato2011,Laigle2016}, whose errors should mostly be smaller than the bin size, so that we can assign them unambiguously into redshift bins. As a result, it seems that the absorbed fraction could be a bit higher at redshift lower than 1, reaching around 0.6. Beyond redshift 1, the absorbed AGN fraction is relatively stable around 0.5.

%__________________________________________________________________
\section{Discussion}
\label{discussion}

\subsection{Bayesian framework of X-ray spectral analysis and population inference}

We have developed a fully Bayesian framework to analyze a large sample of AGN X-ray spectra, from the spectral fitting of each source to the inference of parent distributions of the main spectral parameters, for example\ $N_\mathrm{H}$ and $\Gamma$. Since surveys mostly consist of low-S/N data, the key point of the spectral fitting step was to properly propagate to the parameters of interest the uncertainties associated with the presence of additional, ill-constrained components such as reflection and soft-excess, that are present in the X-ray spectra of AGN, but not always detectable. We used a consistent physical model that includes most of the physical components in AGN to fit the data, irrespective of the S/N, and used informative priors derived from deep surveys to constrain secondary emission components (see Paper~\rom{1} for details). In the end we marginalized over the nuisance parameters and obtained reliable PDFs for the main parameters.

However, despite being more reliable, these PDFs can be loosely constrained and present non-Gaussian shapes, and sometimes even show multi-modality, which makes it very difficult to make correct population inference based on them. The standard population inference method presented in Sect.~\ref{parametric method} is able to make a proper modeling and take into account the full PDFs, but it assumes they always provide the true posterior probabilities (i.e., they are unbiased). We find clear cases this assumption is violated. In particular, the PDFs of $N_\mathrm{H}$ for unabsorbed objects are biased, because the likelihood may become insensitive to the value of the parameter if it is too low, resulting in a flat posterior, limited only by the prior. The bias is a priori unknown, and therefore cannot be easily corrected. In the new method (FFPI) we proposed here, we modeled the stacked PDFs by summing the PDFs that are expected from the parent distribution. These expected PDFs were determined using a transfer matrix, in analogy with the standard way of analyzing X-ray spectra, that takes into account the biases resulting from the fitting process. The matrix is constructed by simulating a large realistic sample of sources that are added onto the real image data and extracted with the same source detection tool and the same background as those of real sources, such that the simulated sources have the same selection function as the real sources (see Sect.~\ref{simulation} for details). 

\begin{figure}[tb]
    \centering
        \includegraphics[width=\columnwidth]{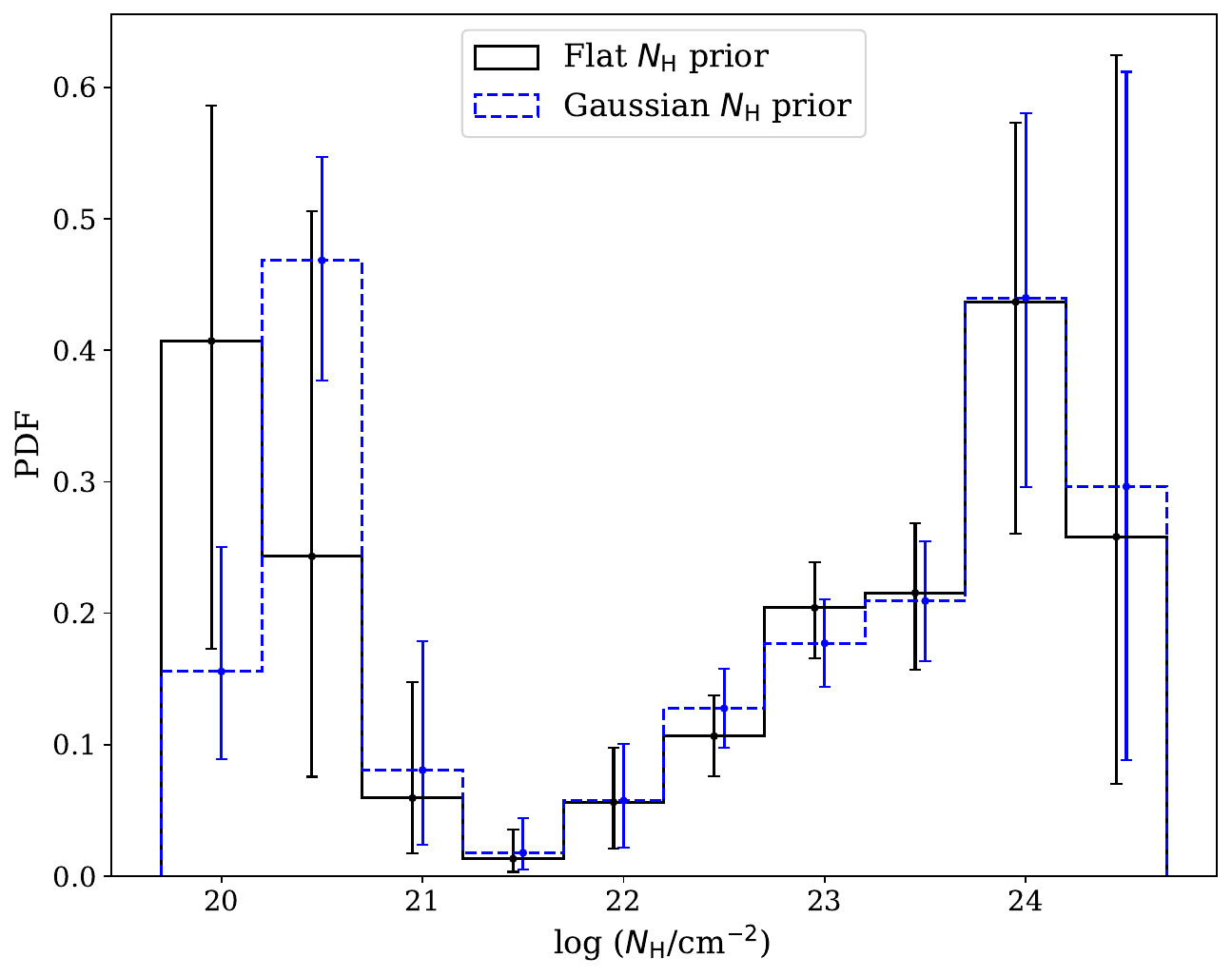}
    \caption{Reconstructed $N_\mathrm{H}$ distributions (selection corrected) of the COSMOS data sample using transfer matrices derived from the full simulated sample with flat-in-log $N_\mathrm{H}$ distribution (solid black), and from the subsample with a log-normal $N_\mathrm{H}$ distribution (average $\log N_\mathrm{H}$ of 22 and dispersion of 1; dashed blue). The error bars show 68.3\% credible intervals. All distributions are normalized to unit area.}
    \label{fig:nh_test}
\end{figure}

To reduce the dimensionality of the problem, we constructed the transfer matrix, for a specific application by marginalizing over the other parameters. One may have the concern that the simulation is based on the assumptions we make on the distributions of these parameters, and thus the transfer matrix may not be representative of the real data. However, just like the priors we used in the Bayesian analysis, these assumptions were also priors that reflect our state of knowledge of this problem, which are consistent with those we used in spectral fitting. In Sect.~\ref{Validation of FFPI}, the relations between the absorbed fraction and redshift and $L_\mathrm{X}$ were successfully recovered for different subsamples with different input relations, using the same transfer matrix computed from the total simulated sample. To further illustrate this point, we selected a subsample with 3000 sources from the full simulated sample, imposing a log-normal distribution with an average $\log N_\mathrm{H}$ of 22 and dispersion of 1, which is practically the opposite of the posterior we found using a flat prior. The reconstructed $N_\mathrm{H}$ distribution of the COSMOS sample using the transfer matrix derived from this sample compared with previous results obtained with the full sample and a flat prior is shown in Fig.~\ref{fig:nh_test}. We find consistent results considering the uncertainties, especially in the range of $10^{21.5}$ to $10^{23.5}$\,cm$^{-2}$, where $N_\mathrm{H}$ is well constrained with our data. This demonstrates that our results are not very sensitive to the prior set by the choice of the simulated subsample we used to construct the transfer matrix, even though we imposed on purpose a very unrealistic prior.

Apart from the priors of the spectral parameters, another prior we put in the simulation is the XLF, which is essential for simulating a sample with realistic $L_\mathrm{X}$ and redshift distribution. However, the parameterization we used does not include the dependence on spectral parameters such as $N_\mathrm{H}$, and hence any difference we found between absorbed and unabsorbed objects cannot be due to the prior. The transfer matrix also takes into account the statistical uncertainty in the output posterior distribution, so that the method naturally deconvolves the stacked PDFs from this effect.

The performance of FFPI in recovering the parent distributions of $\Gamma$, $N_\mathrm{H}$, as well as inferring the redshift and the intrinsic X-ray luminosity dependence of the absorbed fraction, was tested on a realistic simulated data set. We see that using point estimators like the median, or with the simple stacked PDF one cannot recover the true properties of the parent population. For instance, the medians of the PDFs of $N_\mathrm{H}$ indicate a false peak in the unabsorbed regime and another one in the absorbed regime, which is biased toward stronger absorption, whereas the stacked PDF hardly recovers the bi-modality and severely underestimates the peak of absorbed sources (see Fig.~\ref{fig:simul_nh_redist}). However, unlike these two naive approaches, the parametric fitting approach introduced in Sect.~\ref{parametric method} is able to recover the true population, as long as the PDFs are unbiased, for example in the case of $\Gamma$ (see Fig.~\ref{fig:gamma_parametric}), provided we have a reasonable analytic expression for the distribution, which is not always available.

\subsection{Population properties of the AGN X-ray spectral parameters}

We discuss here our results of applying FFPI to \emph{XMM}-COSMOS and compare them with previous studies. We point out however that most other AGN X-ray population studies make strong assumptions about some parameters, for instance, no soft excess or no reflection, which is equivalent to using extremely strong but inappropriate priors, so the comparison with their results must be taken with some caution.

\subsubsection{Photon index $\Gamma$}

We reconstructed the distributions of $\Gamma$ for absorbed and unabsorbed AGN separately for two reasons. First, they are found to be different in some previous works (e.g., \citealt{Burlon2011,Ueda2014}), and second, it is a natural outcome from our method, because we take into account $N_\mathrm{H}$ in the transfer matrix of $\Gamma$. The two distributions were simply marginalized over different $N_\mathrm{H}$ ranges, adopting a flat prior on $N_\mathrm{H}$.

The results we found using FFPI from our \emph{XMM}-COSMOS sample show that the majority of AGN in both absorbed and unabsorbed populations have $\Gamma$ between 1.9 and 2.1, as shown in Fig.~\ref{fig:gamma_parentfit}, which is about 0.2 higher than the typical values found in \citet{Ricci2017} for Swift/BAT AGN, with $\Gamma = 1.80 \pm 0.02$ for unabsorbed AGN and $1.76 \pm 0.02$ for absorbed AGN. The reason why we find on average slightly higher $\Gamma$ values could be that we apply a consistent model, which includes reprocessed emission like reflection, for all the sources, unlike in \citet{Ricci2017}, where secondary components are added on a case-by-case basis and many objects have only an upper limit to the relative reflection. Reflection flattens the slope of the spectrum, because it adds photons in the hard X-ray band. Namely, there is a degeneracy between a single hard power law and a softer one plus reflection. Quantitatively, if we fit a redshifted cutoff power law model (\texttt{zcutoffpl}) to a faked \textit{XMM-Newton} spectrum of a power law plus reflection (\texttt{zcutoffpl} + \texttt{pexmon}) at redshift 0.5 with the relative-reflection parameter $R=1$ (-1, to use the \texttt{xspec} convention to ignore the direct emission in \texttt{pexmon}), we get a fit having a harder photon index by about 0.1. Since we take into account the possibility of the presence of reflection by applying a flat prior between 0 and 1.5 on $R$, which is little constrained in most of our sources, marginalizing over the posterior of $R$ results in larger values of $\Gamma$. On the other hand, \citet{Ueda2014} fixed $R$ to 1 in all cases, and found that the unabsorbed AGN have an average photon index $\Gamma = 1.94 \pm 0.09$, and the absorbed AGN have $\Gamma = 1.84 \pm 0.15$, which are in good agreement with our results, especially those using the parametric fitting.

Alternatively, the discrepancy between the average photon index estimated here and that determined by \citet{Ricci2017} may arise from uncertainties in the effective area calibration of the various X-ray instruments. Indeed, the calibration of \emph{XMM-Newton} and \emph{Chandra} effective areas is known to be inconsistent at the 10--15\% level. When comparing the fitted temperature of galaxy clusters, \citet{Schellenberger2015} showed that for high temperatures the calibration uncertainties can be as large as 20\%, with \emph{Chandra} returning systematically higher temperatures than \emph{XMM-Newton}. To translate the uncertainty into a photon index, we simulated EPIC-pn Bremsstrahlung spectra with temperatures of 8 keV and 10 keV, respectively, and fitted the simulated spectra with a power law. We found that the corresponding uncertainty in the photon index is about 0.2, which may explain the systematic difference observed here. In the majority of cases, \citet{Ricci2017} used soft X-ray spectra from the XRT instrument \citep{Burrows2005} on \emph{Swift}. Using a simultaneous observation campaign of the quasar 3C 273 with several X-ray observatories, \citet{Madsen2017} showed that \emph{Swift}/XRT returns a photon index of 1.5 compared to 1.7 for \emph{XMM-Newton}/EPIC. The harder photon indices measured by \citet{Ricci2017} compared to our work may thus be explained by a difference in the effective area calibration of \emph{XMM-Newton}/EPIC with respect to \emph{Swift}/XRT.

Another difference between our results and those of \citet{Ricci2017} is that we have more constrained distributions for both absorbed and unabsorbed AGN, while the $\Gamma$ distributions in \citet{Ricci2017} are broader, especially for the absorbed AGN. This larger scatter can be explained by the statistical uncertainties on the photon index measurement of each spectrum, which are generally on the order of 0.2--0.3, giving rise to the broadening of the $\Gamma$ distribution. On the other hand, with FFPI, the transfer matrix not only removes the bias of the PDFs, but also the scatter induced by measurement uncertainties, so that the resulting parent distribution is minimally affected by the statistical uncertainties of each measurement. The parametric fitting approach is also able to handle this statistical uncertainty, but not the bias in the PDFs as discussed above.

Although our results for the $\Gamma$ distributions of absorbed and unabsorbed objects are comparable to each other to the first order due to the prominent peak at $\Gamma=2.0$, there is also evidence that the unabsorbed population contains more soft objects than the absorbed one, while the latter has more hard objects (see Fig.~\ref{fig:gamma_parentfit}). Using the parametric fitting we obtained two Gaussian distributions with $\Gamma = 2.05 \pm 0.24$ for the unabsorbed sample and $\Gamma = 1.85 \pm 0.12$ for the absorbed one, showing again that unabsorbed AGN have on average softer photon indices, which is in agreement with \citet{Burlon2011} and \citet{Ueda2014}. The difference might be a consequence of a relationship between Eddington rate ($\lambda_\mathrm{Edd}$) and obscuration \citep{Ricci2017b}, combined with the extensively studied positive correlation between $\Gamma$ and $\lambda_\mathrm{Edd}$ \citep{Brandt1997,Lu1999,Shemmer2006,Shemmer2008,Risaliti2009,Brightman2013,Kawamuro2016,Ricci2018}, which could be explained by a more efficient cooling of the X-ray emitting corona at higher $\lambda_\mathrm{Edd}$ \citep{Vasudevan2007,Davis2011}.

\subsubsection{Absorbing column density, $N_\mathrm{H}$}

We did not attempt to perform a parametric inference for the $N_\mathrm{H}$ distribution. We only used our FFPI forward-fitting inference approach because it is clear from Paper~\rom{1} that the PDFs of $N_\mathrm{H}$ are strongly biased. In addition, it is not clear which functional form should be used to describe the distribution of $N_\mathrm{H}$.

The reconstructed $N_\mathrm{H}$ distribution of our \emph{XMM}-COSMOS sample shown in Fig.~\ref{fig:nh_parentfit} is in good agreement with the results from \citet{Ueda2014} and \citet{Ricci2017}, both of which use local AGN samples detected by Swift/BAT in the 14--195\,keV energy range, which does not suffer too strongly from the incompleteness problem at high column densities. \citet{Ueda2014} utilize the Swift/BAT 9-month sample \citep{Tueller2008}, which contains 137 non-blazar AGN, while \citet{Ricci2017} adopt the extended 70-month sample with 733 non-blazar AGN \citep{Baumgartner2013}, and both of them fit the parameters using complementary observations below 10\,keV.

Besides the good match, we can identify very clearly a bimodality in the $N_\mathrm{H}$ distribution that is only suggested by \citet{Ueda2014} and \citet{Ricci2017}. The two peaks are widely separated, with one peak in the unabsorbed regime around $10^{20}$\,cm$^{-2}$, and the other in the absorbed regime around $10^{23}$ -- $10^{24}$\,cm$^{-2}$. Even though the uncertainties on the two peaks are relatively large, the decreasing trend of the $N_\mathrm{H}$ distribution of unabsorbed sources from $10^{20}$ to $10^{21.5}$\,cm$^{-2}$ and the increasing trend from $10^{21.5}$ to $10^{23}$\,cm$^{-2}$ are very clear. Above $10^{23}$\,cm$^{-2}$, the correction of selection effects starts to be dominant and even shifts the peak from the original distribution, at $10^{23}$\,cm$^{-2}$, to an order of magnitude higher. In view of the large uncertainties, the location of the peak cannot be precisely determined, but it is very likely to be close to $10^{24}$\,cm$^{-2}$. In addition, our results indicate that there is practically no source with $N_\mathrm{H}$ around $10^{21.5\pm0.25}$\,cm$^{-2}$, which is in good agreement with the results of \citet{Merloni2014}, where a boundary at $N_\mathrm{H}=10^{21.5}$\,cm$^{-2}$ is suggested to classify absorbed and unabsorbed AGN. This clear gap is a strong indication that the absorbing materials of these two populations have different origins \citep{Paltani2008}. The one peaked at low $N_\mathrm{H}$ is likely due to the absorption by the host galaxy itself, while the other one is caused by the denser torus around the nucleus.

\subsubsection{Absorbed AGN fraction versus intrinsic X-ray luminosity}

The absorbed AGN fraction is an important indicator in the study of AGN populations, because we can use it to infer the geometry and characteristics of the absorbing material. For example, the simple unification model of AGN \citep{Antonucci1993}, in which the amount of obscuration is only an effect of inclination in a homogeneous population, predicts that the absorbed AGN fraction should be independent of all other physical parameters. However, we find a clear decrease of the absorbed fraction with increasing intrinsic 2--10\,keV X-ray luminosity from around $10^{43}$ to $10^{45}$\,erg\,s$^{-1}$, so this simple model can be ruled out. This dependence was found in previous studies \citep{Hasinger2008,Burlon2011,Ueda2014,Buchner2015}; the results of \citet{Burlon2011} and \citet{Ueda2014} are shown in Fig.~\ref{fig:obsfrac_lx}, which are in broad agreement with our results considering the uncertainties. Similarly, \citet{Ricci2017} also found that unabsorbed AGN have typically higher intrinsic luminosities than absorbed ones. \citet{Ricci2017b} further demonstrated that the dependence disappears when the sources are separated into bins of $\lambda_\mathrm{Edd}$, and thus the main driver of the decreasing absorbed fraction may be $\lambda_\mathrm{Edd}$ instead of $L_\mathrm{X}$.

On the high-$L_\mathrm{X}$ end, our results are slightly above those of \citet{Ueda2014} around $10^{44.5}$ to $10^{45}$\,erg\,s$^{-1}$, and the decreasing trend may flatten above $10^{45}$\,erg\,s$^{-1}$. However, this is very uncertain because of the scarcity of objects. Similarly, on the low-$L_\mathrm{X}$ end, because of the large uncertainties of the first two $L_\mathrm{X}$ bins, our result is still compatible with all published studies, including \citet{Brightman2011}, \citet{Burlon2011} and \citet{Buchner2015}, which found a positive relation when $L_\mathrm{X}$ is below $10^{43}$\,erg\,s$^{-1}$. As shown in Fig.~\ref{fig:obsfrac_lx}, our result between $10^{42}$ to $10^{43.4}$\,erg\,s$^{-1}$ match well with those of \citet{Burlon2011}, especially regarding the location of the peak. More data are needed to better constrain the dependence of the absorbed fraction on $L_\mathrm{X}$ below $10^{43}$\,erg\,s$^{-1}$, which requires surveys of larger area, for example the XXL survey \citep{Pierre2016}, and/or deeper exposure \citep{Giacconi2002}.

\subsubsection{Absorbed AGN fraction versus redshift}

To study the evolution of the absorbed AGN fraction with redshift, we reconstructed the $N_\mathrm{H}$ distribution in each redshift bin, and then calculate the respective absorbed fractions. As shown in Fig.~\ref{fig:obsfrac_z}, to first order we see a flat distribution of the absorbed fraction as a function of redshift for $z\ge 1$, with a slight increase at lower redshifts. We compared our results with those of \citet{Hasinger2008}, which were obtained using a combined sample of ten independent hard X-ray samples selected in the 2--10\,keV band. Their total sample consists of 1290 sources, 1106 of which have reliable optical spectroscopic redshifts. Since their results are not corrected for selection effects, we plot also ours without selection correction in Fig.~\ref{fig:obsfrac_z} for comparison. The results indicate that the overall absorbed fraction from \citet{Hasinger2008} is underestimated because of the selection effects, and even when selecting in the 2--10\,keV range our selection is slightly biased toward  unabsorbed sources. Although the selection functions are different in the two data samples, both results are in relatively good agreement with each other. We measured however overall a slightly lower absorbed fraction at $z>1$ and higher for low-redshift ones, but the difference is within $1\sigma$.

As discussed in \citet{Hasinger2008}, since in each redshift bin the absorbed AGN fraction is integrated over luminosity, its correlation with redshift may be canceled out by the anticorrelation with $L_\mathrm{X}$. It would be better to reconstruct the $N_\mathrm{H}$ distribution in the joint redshift and luminosity space, in order to break the degeneracies. Following a similar approach, \citet{Hasinger2008} obtained a positive correlation between the absorbed fraction and redshift, which was also found in \citet{Ueda2014} and \citet{Buchner2015}. However, in this work the limited size of our COSMOS sample made it unfeasible to perform an accurate three-dimensional reconstruction of the parameter space using FFPI, since the number of sources per three-dimensional bin will be too small to get meaningful measurements. Larger surveys are required to study the evolution of the absorbed AGN fraction with redshift.

 %______________________________________________________________
\section{Summary and conclusions}
\label{conclusions}

In this paper we have presented\ FFPI, a novel nonparametric Bayesian method of population inference that we applied to the reconstruction of the parent distributions of AGN X-ray spectral parameters of interest, for example $\Gamma$, $N_\mathrm{H}$, and $L_\mathrm{X}$, based on the PDFs of the parameters of each individual spectrum in a large survey where most sources have low to moderate S/N.

We tested both a parametric fitting approach and FFPI on simulated data. The parametric approach works well when a parametric form of the parent distribution exists, provided that the PDFs are unbiased (e.g., $\Gamma$; see Sect.~\ref{parametric method}). As some parameters have highly biased PDFs (e.g., $N_\mathrm{H}$, see Sect.~\ref{new method}), we used FFPI to correct the systematic bias in the PDFs by applying a transfer matrix to the stacked PDF of the total sample that models the bias introduced by the spectral fitting process, and by forward-fitting the nonparametric histogram model to the redistributed stacked PDF. The transfer matrix, $M_\mathrm{red}$, is computed from a large realistic simulation that reproduces the redshift and S/N distributions of our sample, on which $M_\mathrm{red}$ is strongly dependent. The sample was simulated using the same priors as those of the spectral parameters applied in the spectral fitting step so the other parameters are marginalized over consistently. The tests on the subsamples of the simulation, which have different parent distributions of $N_\mathrm{H}$, $\Gamma,$ and so on, show that FFPI is able to reconstruct the parent distributions with small biases and reasonable uncertainties, even when using  poorly constrained and sometimes highly biased PDFs (Sect.~\ref{new method}).

As a pilot study for our new Bayesian methods, we applied the spectral analysis method presented in Paper~\rom{1} to the raw \textit{XMM-Newton} observations of the COSMOS field, and reconstructed the parent distributions of the main parameters using FFPI, as well as the parametric approach for the photon index, $\Gamma$. We found a clear bimodality in the $N_\mathrm{H}$ distribution, with widely separated peaks, and an almost total lack of objects with $N_\mathrm{H}$ around $10^{21.5}$\,cm$^{-2}$. The $\Gamma$ distributions of absorbed and unabsorbed AGN are similar, with the common feature that the majority of objects have spectral indices close to 2, which is around 0.1--0.2 higher than the typical values found for Swift/BAT AGN \citep{Ricci2017}. The discrepancy can possibly be explained by calibration issues or by the difference in the spectral modeling. However, the asymmetries in the distributions for absorbed and unabsorbed objects indicate that there are more soft ($\Gamma>2$) objects in the unabsorbed sample than in the absorbed one, and conversely more hard ($\Gamma<2$) objects in the absorbed sample than in the unabsorbed one. We also recovered the well-known anticorrelation between the absorbed AGN fraction and the intrinsic 2--10\,keV $L_\mathrm{X}$ when $L_\mathrm{X}$ is higher than about $10^{43}$\,erg\,s$^{-1}$. Finally, we did not find a clear evolution of the absorbed fraction with redshift: the absorbed fraction is mostly flat and close to 0.5, with a small excess below $z=1$. However, more sources are needed to disentangle the luminosity dependence in each redshift bin.

With the application of our set of Bayesian methods on the \emph{XMM}-COSMOS survey, we have demonstrated that it is possible to reconstruct distributions of the main AGN X-ray spectral parameters that are essentially consistent with previous results, albeit in an automatic way and using a single physically motivated spectral model and a single sample limited to 819 sources with mostly low S/N. Our method can be readily applied to larger surveys, for example the XXL \citep{Pierre2016} and the eROSITA all-sky survey \citep{Merloni2012} surveys, which will allow us to perform a detailed study of the parameters of interest and of their interplay, which is our next step in the application of our methods.

\begin{acknowledgements}
      L.G. acknowledges partial support from the Swiss National Science Foundation. 
\end{acknowledgements}

%-------------------------------------------------------------------
\bibliography{ref.bib}

%\clearpage
\begin{appendix}

%\section{Background estimation}
%\label{Background estimation}
%
%We use spectra extracted from an annulus 60 to 90 arcsec away from the source center to infer the true background spectra in the source region (Sect.~\ref{source detection}), assuming there is no background variation and contamination from different sources. With the simulated sample we can actually see how good the background estimated in this way is, since we know the true background of the simulated sources.
%
%In Fig.~\ref{fig:bg_estimate} we show the scatter of the estimated background count rates versus the true ones for both detectors in the 2--7\,keV detection band. The background estimation is mostly constrained within 15\% uncertainties without noticeable bias for both pn and MOS. However, there are quite expectedly a few outliers that are the cause of most of the spectra with negative S/N shown in Fig.~\ref{fig:sample_cr_snr}.
%
%\begin{figure}[tb]
%\centering
%\includegraphics[width=\columnwidth]{pictures/simulation/bg_estimate.pdf}
%\caption{True background count rates vs those estimated from background regions around the sources of the simulated sample. MOS and pn count rates in the 2--7\,keV source detection band are plotted in blue and pink. The dashed line is the 1:1 relation, and the dotted lines correspond to 15 percent deviation from the reference line.}
%\label{fig:bg_estimate}
%\end{figure}

\section{Photon smearing induced by the \emph{XMM-Newton} PSF}
\label{sec:psf}

While our choice of source extraction region (a circle with a radius of 30 arcsec) is large enough to encompass the majority of source photons, a fraction of the incoming photons will be scattered by the \emph{XMM-Newton} PSF outside of the region of interest, such that the measured count rates will be slightly underestimated. For simplicity, in this paper we do not attempt to correct for this effect. Here we verify that the loss of photons induced by PSF smearing does not significantly bias our measurements.

\begin{figure}
    \centering
    \resizebox{\hsize}{!}{\includegraphics{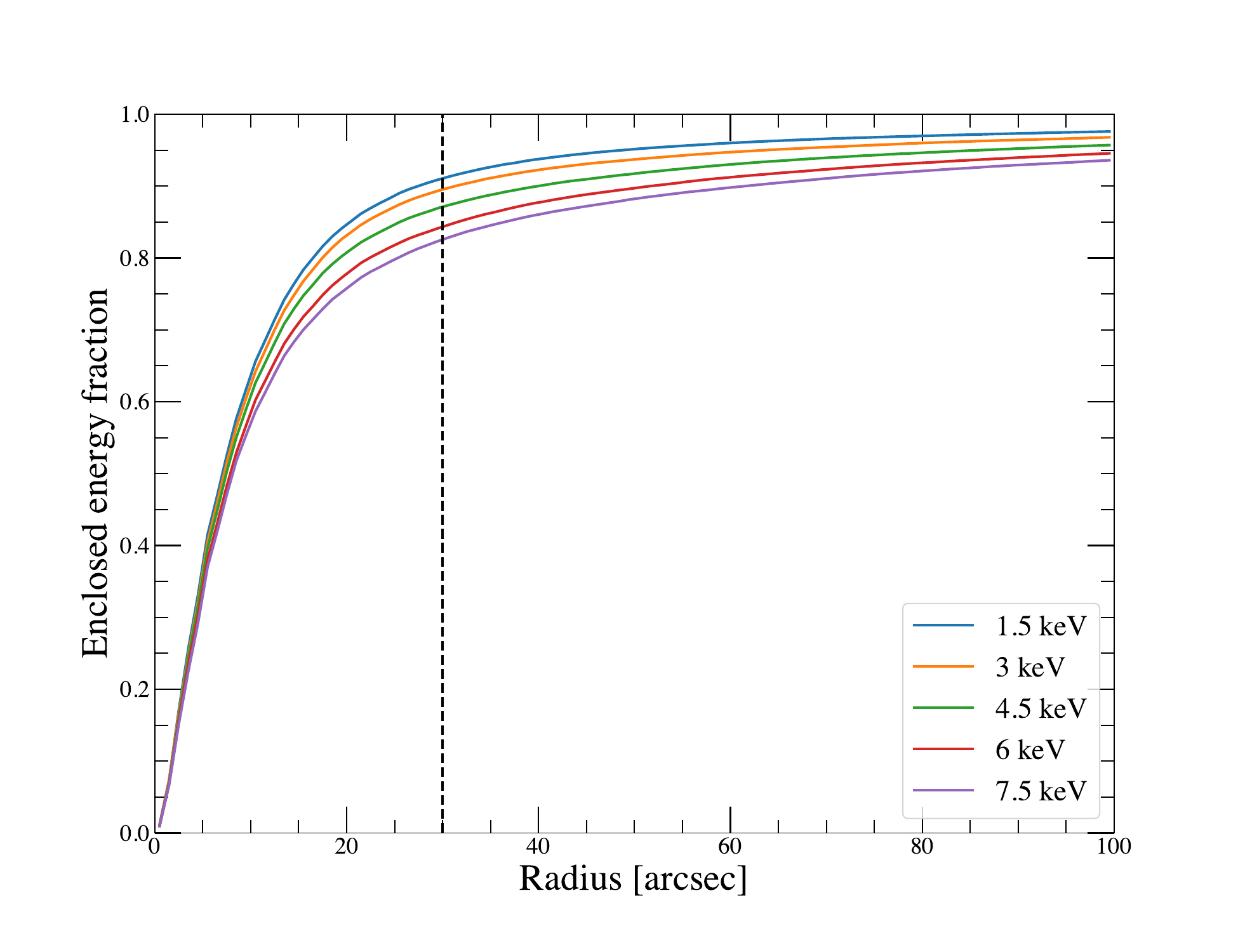}}
    \caption{Enclosed energy fraction of the \emph{XMM-Newton} PSF at 8 arcmin off-axis angle as a function of distance to the source position for several energies. The dashed vertical line shows the radius of 30 arcsec used as our source extraction radius.}
    \label{fig:psf_enc}
\end{figure}

The \emph{XMM-Newton} PSF has a half-energy width of 8 arcsec diameter on-axis at 1 keV. However, the half-energy width degrades with off-axis angle and energy, such that the enclosed energy fraction is not uniform. To verify the impact of PSF smearing on our results, we use the model PSF images available in the official \emph{XMM-Newton} calibration database \citep{Read2011} to compute the enclosed fraction within a circle of 30 arcsec radius as a function of energy, for a FOV-averaged off-axis angle of 8 arcmin. In Fig. \ref{fig:psf_enc} we show the enclosed energy fraction as a function of distance to the source position for several energies. The energies below 1.5 keV are not shown as the PSF does not significantly depend on energy below that point. We can see that the energy enclosed within a circle of 30 arcsec radius slightly decreases with energy, from 91\% at 1.5 keV to 82\% at 7.5 keV. Therefore, neglecting the impact of photon loss affects the measurement of the spectral shape.

To quantify the impact of photon loss on the recovered spectral parameters, we performed an EPIC-pn on-axis simulation of a high S/N source with an unabsorbed power law spectrum with a typical photon index of 1.9. We then modified the simulated spectrum according to the enclosed energy fraction shown in Fig. \ref{fig:psf_enc}, reducing the count rate at each energy by the corresponding encircled energy. We then fitted the modified spectrum with an unabsorbed power law and checked how the best-fit parameters differ from the input ones. We found that the photon index is essentially unaffected by this procedure, as the output photon index is mainly driven by the shape of the low-energy data ($<2$ keV), which dominate the photon statistic. Conversely, the fitted normalization of the spectrum is reduced by 10\% compared to the true normalization. We obtain similar results when considering an absorbed power law with an absorbing column density of $\log N_H=22.5$. Therefore, we conclude that our measurements of $\Gamma$ and $N_H$ are unaffected by the photon loss. However, our recovered luminosities are systematically underestimated by $10\%$, which can be ignored for the sake of this paper. Future versions of our code will include a correction for photon loss as a function of energy to avoid biasing our fitted luminosities toward low values.

\section{Observation list}
\label{observation list}

\begin{table*}[!b]
\centering 
\caption{List of observations.}
\begin{tabular}{c c c c c c c c}        
\hline\hline
\rule{0pt}{1.2em}\noindent
ObsId & COSMOS field \# & RA & Dec \\  
\hline
\rule{0pt}{1.2em}\noindent
   0203360101  &  1    &   150.61005  &   2.71  \\
   0203360201  &  2    &   150.61005  &   2.46  \\
   0203360301  &  3    &   150.61005  &   2.21  \\
   0203360401  &  4    &   150.61005  &   1.96  \\
   0203360501  &  5    &   150.61005  &   1.71  \\
   0203360601  &  6    &   150.36000  &   2.71  \\
   0203360701  &  7    &   150.36000  &   2.46  \\
   0203360801  &  8    &   150.36000  &   2.21  \\
   0203360901  &  9    &   150.36000  &   1.96  \\
   0203361001  &  10   &   150.36000  &   1.71  \\
   0203361101  &  11   &   150.10995  &   2.71  \\
   0203361201  &  12   &   150.10995  &   2.46  \\
   0203361301  &  13   &   150.10995  &   2.21  \\
   0203361401  &  14   &   150.10995  &   1.96  \\
   0203361501  &  15   &   150.10995  &   1.71  \\
   0203361601  &  16   &   149.85999  &   2.71  \\
   0203361701  &  17   &   149.85999  &   2.46  \\
   0203361801  &  18   &   149.85999  &   2.21  \\
   0203361901  &  19   &   149.85999  &   1.96  \\
   0203362001  &  20   &   149.85999  &   1.71  \\
   0203362101  &  21   &   149.61000  &   2.71  \\
   0203362201  &  22   &   149.61000  &   2.46  \\
   0203362301  &  23   &   149.61000  &   2.21  \\
   0203362401  &  24   &   149.61000  &   1.96  \\
   0203362501  &  25   &   149.61000  &   1.71  \\
   0302350101  &  B1   &   150.61005  &   2.73  \\
   0302350201  &  B2   &   150.62670  &   2.46  \\
   0302350301  &  B3   &   150.61005  &   2.23  \\
   0302350401  &  B4   &   150.62670  &   1.96  \\
   0302350501  &  B5   &   150.61005  &   1.73  \\
   0302350601  &  B6   &   150.34335  &   2.71  \\
   0302350701  &  B7   &   150.36000  &   2.48  \\
   0302350801  &  B8   &   150.34335  &   2.21  \\
   0302350901  &  B9   &   150.36000  &   1.98  \\
   0302351001  &  B10  &   150.34335  &   1.71  \\
   0302351101  &  B11  &   150.10995  &   2.73  \\
   0302351201  &  B12  &   150.12660  &   2.46  \\
   0302351301  &  B13  &   150.10995  &   2.23  \\
   0302351401  &  B14  &   150.12660  &   1.96  \\
   0302351501  &  B15  &   150.10995  &   1.73  \\
   0302351601  &  B16  &   149.84333  &   2.71  \\
   0302351701  &  B17  &   149.85999  &   2.48  \\
   0302351801  &  B18  &   149.84333  &   2.21  \\
   0302351901  &  B19  &   149.85999  &   1.98  \\
   0302352001  &  B20  &   149.84333  &   1.71  \\
   0302352201  &  B22  &   149.62667  &   2.46  \\
   0302352301  &  B23  &   149.61000  &   2.23  \\
   0302352401  &  B24  &   149.62667  &   1.96  \\
   0302352501  &  B25  &   149.61000  &   1.73  \\
   0302353001  &  B9   &   150.36000  &   1.98  \\
   0302353101  &  B3   &   150.61005  &   2.23  \\
   0302353201  &  B24  &   149.62667  &   1.96  \\
   0302353301  &  B25  &   149.61000  &   1.73  \\
   0302353401  &  B14  &   150.12660  &   1.96  \\
   0501170101  &  20C  &   149.84333  &   1.71  \\
   0501170201  &  23C  &   149.61000  &   2.23  \\
\hline\hline                                  
\end{tabular}
%\tablefoot{Column (1).}
\label{obslist}
\end{table*}

\end{appendix}

\end{document}